\documentclass[11pt]{article}

\usepackage{array}
\newcolumntype{P}[1]{>{\centering\arraybackslash}p{#1}}

\usepackage{booktabs} % move this to preamble and uncomment

\usepackage{hyperref}

\usepackage{tikz}
\usepackage{verbatim}

\usepackage{amssymb}
\usepackage{import}
\usepackage{lineno}
\usepackage{mathtools,bbm}
\usepackage{graphicx,url}
\usepackage{graphics}

\DeclareMathOperator*{\E}{\mathbb{E}}
\DeclareMathOperator*{\R}{\mathcal{R}}

\DeclareMathOperator*{\I}{\mathbbm{1}}
\DeclareMathOperator*{\F}{\mathcal{F}}
\DeclareMathOperator*{\X}{\mathcal{X}}

\DeclareMathOperator*{\argmin}{argmin}

\usepackage{setspace,graphicx,epstopdf,amsmath,amsfonts,amssymb,amsthm,versionPO}
\usepackage{marginnote,datetime,enumitem,rotating,fancyvrb,scalerel}
\usepackage{hyperref}
\usepackage{pdfpages}

\excludeversion{notes}		% Include notes?
\includeversion{links}          % Turn hyperlinks on?

\iflinks{}{\hypersetup{draft=true}}

\ifnotes{%
\usepackage[margin=1in,paperwidth=10in,right=2.5in]{geometry}%
\usepackage[textwidth=1.4in,shadow,colorinlistoftodos]{todonotes}%
}{%
\usepackage[margin=1in]{geometry}%
\usepackage[disable]{todonotes}%
}

\makeatletter\let\chapter\@undefined\makeatother % Undefine \chapter for todonotes

\usepackage{indentfirst} % Indent first sentence of a new section.
\usepackage{jfe}          % JFE-specific formatting of sections, etc.

\newtheorem{proposition}{Proposition}

\newtheorem{lemma}{Lemma}[section]

\newtheorem{definition}{Definition}

\usepackage[
  backend=biber,
  style=numeric, % default
]{biblatex}
\usepackage{float}
\usepackage[caption = false]{subfig}

\begin{document}

\setlist{noitemsep}  % Reduce space between list items (itemize, enumerate, etc.)
%\onehalfspacing      % Use 1.5 spacing
% Use endnotes instead of footnotes - redefine \footnote command

\title{\textbf{Deep Neural Networks for Choice Analysis: A Statistical Learning Theory Perspective}}

\author{Shenhao Wang \\
  Qingyi Wang \\
  Nate Bailey \\
  Jinhua Zhao \\
  \\
  Massachusetts Institute of Technology} 
\date{August 2019}              % No date for final submission

% Create title page with no page number
\renewcommand{\thefootnote}{\fnsymbol{footnote}}

\singlespacing

\maketitle

\vspace{-.2in}
\begin{abstract}
\noindent
While researchers increasingly use deep neural networks (DNN) to analyze individual choices, overfitting and interpretability issues remain as obstacles in theory and practice. By using statistical learning theory, this study presents a framework to examine the tradeoff between estimation and approximation errors, and between prediction and interpretation losses. It operationalizes the DNN interpretability in the choice analysis by formulating the metrics of interpretation loss as the difference between true and estimated choice probability functions. The interpretation of DNN-based choice models relies on function estimation and automatic utility specification, contrary to that of traditional choice models relying on parameter estimation and handcrafted utility specification. This study also uses the statistical learning theory to upper bound the estimation error of both prediction and interpretation losses in DNN, shedding light on why DNN does not have the overfitting issue. Three scenarios are then simulated to compare DNN to binary logit model (BNL). We found that DNN outperforms BNL in terms of both prediction and interpretation for most of the scenarios, and larger sample size unleashes the predictive power of DNN but not BNL. DNN is also used to analyze the choice of trip purposes and travel modes based on the National Household Travel Survey 2017 (NHTS2017) dataset. These experiments indicate that DNN can be used for choice analysis beyond the current practice of demand forecasting because it has the inherent utility interpretation, the flexibility of accommodating various information formats, and the power of automatically learning utility specification. DNN is both more predictive and interpretable than BNL unless the modelers have complete knowledge about the choice task, and the sample size is small ($< 10^4$). Overall, statistical learning theory can be a foundation for future studies in the non-asymptotic data regime or using high-dimensional statistical models in choice analysis, and the experiments show the feasibility and effectiveness of DNN for its wide applications to policy and behavioral analysis. 
\end{abstract}

\medskip

%\noindent \textit{JEL classification}: XXX, YYY.
%\medskip
%\textit{Keywords}: \LaTeX; papers with no content.

\thispagestyle{empty}

\clearpage

\onehalfspacing
\setcounter{footnote}{0}
\renewcommand{\thefootnote}{\arabic{footnote}}
\setcounter{page}{1}

\section{Introduction}
\noindent
Choice modeling is a rich theoretical field widely applied throughout transportation research, and in many other contexts \cite{Train1980,Ben_Akiva1985,Train2009}. While traditional discrete choice models have been used for decades, researchers have recently become increasingly interested in instead using machine learning classifiers to conduct choice analysis due to the high performance of these models in many fields \cite{Karlaftis2011,Paredes2017,Hagenauer2017}. 

Traditional discrete choice models rely on researchers' use of domain knowledge to filter through several model specifications and find the ones that best fit observed data. Machine learning classifiers can improve upon this approach owing to their automated exploration and extraordinary approximation power. By using flexible model family assumptions, the approximation power of many machine learning methods is much higher than discrete choice models, which are typically limited to a linear-in-parameter form with handcrafted features (e.g. quadratic or log forms). Among all machine learning classifiers, the deep neural network (DNN) is particularly powerful due to several factors. It has high approximation power \cite{Hornik1989,Hornik1991,Cybenko1989}, can flexibly accommodate various types of information \cite{Krizhevsky2012,LeCun2015}, has high predictive power as revealed in experimental studies \cite{Fernandez2014,Karlaftis2011}, and has been applied to numerous domains \cite{LeCun2015,Goodfellow2016,Glaeser2018}. However, two unresolved issues hinder the applicability of DNN in many transportation choice analysis contexts: model overfitting in relatively small data sets, and lack of interpretability.

The first concern in applying DNN to transportation choice analysis research is its potential to overfit models to the small data sets typically available in this field. An overfitted model fits the training data precisely but has poor out-of-sample performance. Classical statistical theory suggests that the Vapnik-Chervonenkis (VC) dimension, a measure of model complexity, must be asymptotically small relative to the sample size in order to avoid overfitting \cite{Vapnik1999,Vapnik2013}. However, the scenario of using DNN is typically in a non-asymptotic regime where the classical asymptotic assumption does not hold \cite{Wainwright2019}. While an increasing number of transportation studies use DNN to predict travel choices with high accuracy even on small data sets \cite{Karlaftis2011,Hagenauer2017,Cantarella2005,Dong2018,Mozolin2000,Polson2017,WuYuankai2018}, this theoretical issue remains unresolved and there exist no practical guidelines as to what circumstances may result in overfitting issues when using DNN for choice analysis.

The second concern in transportation choice applications of DNN is its perceived lack of interpretability. Prediction is a typical focus of all modeling whether done via discrete choice models or machine learning classifiers, but many transportation applications require interpretation as well. Interpretability is important for researchers, who seek to understand findings on mode shares, elasticities, marginal rates of substitution, and social welfare, as well as the general public, among whom interpretability has been found useful in building trust \cite{Lipton2016} and explaining results to users \cite{Boshi_Velez2017}. DNN is typically framed as a ``black box'' model, and it is ranked as the model with lowest interpretability among all machine learning classifiers \cite{Kotsiantis2007,Lipton2016,ZhouBolei2014}. The majority of previous studies using DNN for transportation choice modeling have focused narrowly on using DNN to predict mode choice, activity choice, car ownership, or other individual choices \cite{Hensher2000,XieChi2003,Cantarella2005,Celikoglu2006,Omrani2015,Hagenauer2017}. Only a small number of transportation studies touched upon the interpretability of DNN in choice modeling, but do not provide explicit metrics to measure the quality of interpretability \cite{Rao1998,Bentz2000,Hagenauer2017}. The interpretability of DNN models, particularly in comparison to discrete choice models, will be a key factor in determining whether these approaches can be extended to transportation contexts beyond demand prediction and have practical implications on our understanding of individual decision-making behavior.

This paper seeks to address both of these issues through the development of a statistical learning theoretical framework consisting of two dimensions. The first dimension is the decomposition of estimation and approximation errors. We demonstrate that the estimation error of DNN architectures used in choice models is not very large, addressing the first overfitting issue. We present a proof that illustrates that the magnitude of the parameters in DNN is more important than the number of parameters in upper bounding the estimation error in a non-asymptotic way. The second dimension concerns prediction and interpretation losses. Particularly, we substantiate the concept of interpretation by formulating metrics to measure interpretation loss as a counterpart to prediction loss. With our formulation, interpretation loss is measured by the difference between the true and the estimated choice probability functions, drawing on the fact that all valuable economic information can be derived from this function. Model interpretability in DNN relies on the full choice probability function based on automatically learned utility specification. This is in sharp contrast to traditional choice models, which are interpreted through the individual parameters chosen for the utility functions. Through this new conceptualization of interpretation loss, we can evaluate models in terms of both prediction and interpretation losses, allowing us to evaluate and demonstrate the potential for DNN to serve as a powerful predictive and interpretable tool for choice analysis research. 

To illustrate this theoretical framework, we compare DNN to binary logit model (BNL), a representative discrete choice modeling approach through four experiments. Three of these experiments use synthetic data in combination with Monte Carlo simulation, illustrating the tradeoffs between approximation and estimation errors as well as those between interpretation and prediction losses under different sample sizes and input dimensions. The last experiment uses data from the National Household Travel Survey 2017 (NHTS 2017) in order to shed light on the practical relevance of this new theoretical framework, allowing us to provide practical suggestions for future DNN applications in choice modeling research. In all of these experiments, BNL is chosen for comparison with DNN because BNL is more similar to DNN than other DCMs, such as nested and mixed logit models. Whereas BNL is only one member of DCMs, the findings of this paper can also be extended to other DCMs that are not used for comparison in this paper.

This study is the first that introduces a unified framework of statistical learning theory for DNN-based choice analysis. This framework fits the non-asymptotic data regime and forms the foundation for choice modeling with high-dimensional data, which cannot be adequately analyzed by classical statistical tools. We respond to the two critical issues in DNN: overfitting and interpretability, which can guide the process of designing experiments, conducting surveys, training models, and providing policy suggestions in the DNN-based choice modeling. Our experiments elaborate when and why DNN performs better than classical multinomial logit models, with specific modeling suggestions for future studies. The theory and experiments in this paper illustrate the predictive and interpretable potential of DNN models, and demonstrate their use beyond demand forecasting and in domains typically reserved for discrete choice models, such as policy and behavioral analysis.

The paper is organized as follows. In section 2, we describe in more detail the theoretical background and relevant past studies for our framework. In this section, we formulate evaluation metrics for interpretation which can be used for both DCM and DNN and then use statistical learning theory to characterize the four quadrants resulting from the dual tradeoffs between approximation error and estimation error and between prediction and interpretation losses. The introduction of each quadrant is followed by the review of the previous studies most relevant to them. In section 3, we describe our three simulation experiments on synthetic data, illustrating the dynamics of the tradeoffs between the four quadrants. Then we apply our framework to the NHTS data and discuss the resulting findings. Section 4 concludes the paper with remarks on implications and future research.

\section{Theory and Literature Review}
\subsection{Setup of DNN-Based Choice Modeling with Statistical Learning Theory}
\noindent
Let $s(x_i)$ denote the probability of individual $i$ choosing alternative $1$ out of $\{0,1\}$ alternatives, and $x_i$ the inputs including alternative- and individual-specific variables: $s(x_i): R^d \rightarrow [0,1]$. Individual choice $y_i \in \{0,1 \}$ is a Bernoulli random variable with $s(x_i)$ probability of choosing the alternative $1$. This soft decision rule is a common assumption in choice analysis, and it is more generic than the hard decision rule that does not involve probabilistic decisions \footnote{An asymptotically soft decision rule with Softmax or Sigmoid activation function becomes a hard decision rule.}. Let $f(x_i): R^d \rightarrow \{0,1\}$ represent the hard decision rule mapping. Let ${\F}_1$ denote the model class represented by a feedforward DNN, with the layer-by-layer feature transformation $\Phi_1(x_i, w) = (g_m \circ ... g_2 \circ g_1)(x_i)$, in which $g_j(x) = ReLU(\langle W_j, x \rangle)$ representing one standard module in the DNN consisting of ReLU activation and linear transformation. When DNN is applied to a binary choice case, the choice probability $s(x_i, w)$ becomes
\begin{flalign} \label{eq:choice_probability_dnn}
s(x_i, w) = \sigma(\Phi_1(x_i, w)) = \frac{1}{1 + e^{-\Phi_1(x_i, w)}}
\end{flalign}

\noindent
where $\sigma$ is the Sigmoid activation function, and $w$ represents all the coefficients in the DNN. Note that $\Phi_1$ is similar to the deterministic utility difference $V_1 - V_0$ in choice models. With larger $\Phi_1$, individual $i$ is more likely to choose alternative $1$ over $0$. Let ${\F}_0$ represent the model class of BNL and $\Phi_0(x_i, w) = \langle w, x_i \rangle$ represent the linear feature mapping in BNL. It can be shown that BNL is a special case of DNN (shown in Appendix I): ${\F}_0 \subset {\F}_1$. The choice probability of $s(x_i)$ in BNL is similar to Equation \ref{eq:choice_probability_dnn}, except for replacing $\Phi_1$ with $\Phi_0$. Let $S = \{x_i, y_i \}_{i=1}^N$ denote the sample; $N$ the sample size; $x \sim P_x(x)$ the data generating process of $x$; and $s^*(x)$, $f^*(x)$, and $w^*$ the true models and parameters. Empirical risk minimization is used to obtain their estimators: $\hat{s}(x)$, $\hat{f}(x)$, and $\hat{w}$.

\begin{definition} \label{def:erm}
Empirical risk minimization (ERM) is defined as
\begin{flalign}
\underset{f \in \F }{\min} \ \hat L(f) = \underset{f \in \F }{\min} \ \frac{1}{N} \sum_{i=1}^N l(y_i, f(x_i))
\end{flalign}
Estimator based on ERM is defined as
\begin{flalign}
\hat{f} = \underset{f \in \F }{\argmin} \ \hat L(f) = \underset{f \in \F }{\argmin} \ \frac{1}{N} \sum_{i=1}^N l(y_i, f(x_i))
\end{flalign}
\end{definition}

\noindent
In training ERM, it is critical to choose a specific \textit{expected loss function} $L(y, x) = \E_{x,y} [l(y, f(x))]$. One common choice is log-loss, which is associated with classical maximum likelihood estimation. To understand the out-of-sample performance of any estimator, we need to examine the \textit{excess error}:

\begin{definition} \label{def:excess_error}
Excess error of $\hat{f}$ is defined as
\begin{flalign}
{\E}_S [L(\hat{f}) - L(f^*)]
\end{flalign}
that of $\hat{s}$ is defined as 
\begin{flalign}
{\E}_S [L(\hat{s}) - L(s^*)]
\end{flalign}
\end{definition}

\noindent
$L(\hat{f})$ and $L(\hat{s})$ are the population error of the estimator, while $L(f^*)$ and $L(s^*)$ are the population error of the true model. Excess error measures to what extent the error of the estimator deviates from the true model, averaged over random sampling $S$. A tight upper bound of excess error can guarantee reliable out-of-sample performance. In the following discussions, we will mainly use $f^*$ and $\hat{f}$ as the running examples, but all the following arguments apply to $s^*$ and $\hat{s}$. Excess error can be decomposed into estimation error and approximation error as following.
\begin{flalign}
{\E}_S [L(\hat f) - L(f^*)] &= {\E}_S[L(\hat f) - L(f^*_F)] + {\E}_S[L(f^*_F) - L(f^*)]
\label{eq:excess_error_decomposition}
\end{flalign}

\noindent
where $f^*_{F} = \underset{f \in \F}{\argmin} \ L(f)$, the best function in function class $\F$ to approximate $f^*$.

\begin{definition} \label{def:estimation_error}
Estimation error refers to 
\begin{flalign}
{\E}_S[L(\hat f) - L(f^*_F)]
\end{flalign}
\end{definition}

\begin{definition} \label{def:approximation_error}
Approximation error refers to 
\begin{flalign}
{\E}_S[L(f^*_F) - L(f^*)]
\end{flalign}
\end{definition}

\noindent
Estimation error is the first term in Equation \ref{eq:excess_error_decomposition} and approximation error is the second term. Estimation error is a quantity that measures whether $\hat{f}$ overfits: very large $L(\hat f) - L(f^*_F)$ implies serious overfitting. Since the estimation error has a $\hat{f}$ term, it captures the randomness from sampling and training. Approximation error is more deterministic and captures only the difference between the best function $f^*_F$ in function class $\F$ and the true function $f^*$. The following four subsections will introduce the prediction loss, the interpretation loss, the approximation error, and the estimation error of DNNs in order. 

\subsection{Prediction Loss}
\begin{definition} \label{def:pred_error}
Prediction loss is defined as
\begin{flalign}
L_{0/1}(f) = {\E}_{x,y} [\I \{ y \neq f(x) \}]
\end{flalign}
Empirical prediction loss is defined as 
\begin{flalign}
\hat L_{0/1}(f) = \frac{1}{N} \sum_{i=1}^N \I \{ y_i \neq f(x_i) \} 
\end{flalign}
\end{definition}

\noindent
Prediction loss is undoubtedly the most common and widely used metric to evaluate prediction performance. Nearly all studies that have used machine learning classifiers to predict any travel-related decision evaluated their models based on the out-of-sample prediction loss \cite{ChengLong2019,Tang2015,Paredes2017,Allahviranloo2013,Hagenauer2017,Cantarella2005,Hensher2000}. The practice of using prediction loss as an evaluation metric also dominates other fields that apply machine learning classifiers to solve practical questions \cite{Krizhevsky2012,LeCun2015,HeKaiming2016}. Several empirical benchmark papers have used prediction loss as the evaluation metric to compare performance across hundreds of models and datasets, thus providing generalizable conclusions \cite{Fernandez2014,Kotsiantis2007}. Our study will also use this prediction loss to evaluate the models for their predictive performance.

\subsection{Interpretation Loss}
\noindent
\begin{definition} \label{def:fun_est_error}
Interpretation loss is defined as the difference between true and estimated choice probability functions
\begin{flalign}
L_e(s) = ||s^* - s||^2_{L^2(P_x)} = \int_{x} (s^*(x) - s(x))^2 d P(x)
\end{flalign}
Empirical interpretation loss is defined as
\begin{flalign}
\hat{L}_s(s) = \frac{1}{N} \sum_{i=1}^N (s^*(x_i) - s(x_i))^2
\end{flalign}
\end{definition}

Interpretation loss is measured by the difference between true and estimated choice probability, integrated over domain $\mathcal{X}$ and weighted by $P_x(x)$. We choose to use this measurement because researchers can obtain most important economic information through the choice probability function $s(x)$. For example, the probability derivatives of choosing alternative $1$ with respect to price $x_j$ can be computed as the derivative $\frac{ds(x)}{dx_j}$; its associated elasticity is $\frac{d \log s(x)}{d \log x_j}$; value of travel time savings (VTTS) can be computed as ratio of two derivatives $\frac{ds(x)/dx_{j1}}{ds(x)/dx_{j2}}$; the utility difference can be computed by using inversed Sigmoid function $V_1 - V_0 = \sigma^{-1}(s)$; or the empirical market share of alternative $1$ can be computed by $\sum_{i=1}^{N} s(x_i)$. Therefore, an accurate function estimator $\hat{s}(x)$ could help recover elasticity values, marginal rate of substitution (such as VTTS), market share, utility values, and social welfare, which provide most of the economic information needed in practice.

It is crucial to see that we focus on \textit{function} estimation $\hat{s}(x)$ rather than \textit{parameter} estimation $\hat{w}$, which is the traditional focus of the majority of the econometric models. The focus on parameter estimation is nearly impossible for DNN for at least three reasons. First, a simple feedforward DNN could easily have tens of thousands parameters, and this large number renders it impossible for researchers to discuss individual parameters. Second, DNN has the property called symmetry of parameter space \cite{Bishop2006}, implying that different parameters could lead to the same choice probability function $s(x)$. Therefore, interpreting individual parameters $w$ is vacuous in DNN. Third, studies have shown that semantic information cannot be revealed from individual neurons, but from the space of each layer in DNN \cite{Szegedy2014}. A large number of studies used the function estimators in DNN for interpretation, while none used individual neurons/parameters \cite{Montavon2018,Hinton2015,Baehrens2010,Ross2018}. Mullainathan and Spiess \cite{Mullainathan2017} argued that ML classifiers (including DNN) are categorically different from econometric models since the ML classifiers focus on $\hat{y}$ while the econometric models focus on $\hat{w}$. This is generally true; however, in the case of DNN, an accurate estimator of the choice probability function $\hat{s}(x)$ could satisfy most of our interpretation purposes traditionally achieved through using $\hat{w}$. In fact, several studies in the transportation field have visualized or computed the gradient information of the choice probability functions to interpret the ML classifiers, supporting our definition of the interpretation loss based on the choice probability functions \cite{Rao1998,Bentz2000,Hagenauer2017}. Moreover, the process of interpretting elasticity $\frac{ds}{dx_j}$ is the same as the discussion of using input gradients in the ML community \cite{Baehrens2010,Montavon2018}. Therefore, the shifting focus from parameter to function estimation enables researchers to interpret DNN results in choice analysis context, and this shift is both inevitable and desirable.

Whereas our definition of the interpretation loss captures the key economic information through the choice probability function, it is not the only way to define interpretation loss. Lipton (2016) \cite{Lipton2016} discussed multiple aspects of interpretability, including simulatability, decomposability, algorithmic transparency, and post-hoc interpretability. Our definition of interpretation loss is focused on the post-hoc interpretability restricted to only economic information, and does not address the other aspects of interpretability and other types of information obtained by using post-hoc interpretation methods \cite{Ribeiro2016,Montavon2018,Hinton2015}. Whereas our approach aligns with the long tradition of choice modeling, it is possible to define interpretation loss in other ways, as shown in a very recent working paper by Bertsimas et al. (2019) \cite{Bertsimas2019}.

\subsection{Approximation Error}
\begin{figure}[ht]
\centering
\subfloat[F0 One-Layer Sparse NN (BNL)]{\resizebox{0.3\linewidth}{!}{% visualize neural net
\def\layersep{3.5cm}
\centering
\begin{tikzpicture}[shorten >=1pt, ->, draw=black!50, node distance=\layersep]
    \tikzstyle{every pin edge}=[<-, shorten <=1pt]
    \tikzstyle{neuron}=[circle,fill=black!25,minimum size=17pt,inner sep=0pt]
    \tikzstyle{input neuron}=[neuron, fill=black!50];
    \tikzstyle{output neuron}=[neuron, fill=red!50];
    \tikzstyle{hidden neuron}=[neuron, fill=blue!50];
    \tikzstyle{annot} = [text width=3cm, text centered]

    % Draw the input layer nodes
%    \foreach \name / \y in {1,...,4}
    % This is the same as writing \foreach \name / \y in {1/1,2/2,3/3,4/4}
    \node[input neuron, pin=left:X0] (X0-0) at (0 cm, 0 cm) {};
    \filldraw [black!50] (0, -0.5cm) circle (2pt);
    \filldraw [black!50] (0, -1.0cm) circle (2pt);
    \node[input neuron] (X0-1) at (0 cm,-1.5cm) {};

    \node[input neuron, pin=left:X1] (X1-0) at (0 cm, -2.5cm) {};
    \filldraw [black!50] (0, -3.0cm) circle (2pt);
    \filldraw [black!50] (0, -3.5cm) circle (2pt);
    \node[input neuron] (X1-1) at (0 cm,-4.0cm) {};

    \node[input neuron, pin=left:Z] (Z-0) at (0 cm, -5.0cm) {};
    \filldraw [black!50] (0, -5.5cm) circle (2pt);
    \filldraw [black!50] (0, -6.0cm) circle (2pt);
    \node[input neuron] (Z-1) at (0 cm,-6.5cm) {};

    % Draw the output layer node
    \foreach \name / \y in {0,...,1}
        \node[output neuron, pin={[pin edge={->}]right:Y \y}] (Y-\name) at (\layersep, -\y cm - 2.5cm) {};

    % Connect every node in the hidden layer with the output layer
    \foreach \source in {0,...,1}
        \foreach \dest in {0}
            \path (X0-\source) edge (Y-\dest);

    \foreach \source in {0,...,1}
        \foreach \dest in {1}
            \path (X1-\source) edge (Y-\dest);

    \foreach \source in {0,...,1}
        \foreach \dest in {1}
            \path (Z-\source) edge (Y-\dest);

%    \foreach \source in {1,...,5}
%        \path (H-\source) edge (O);

    % Annotate the layers
%    \node[annot,left of=hl] {Input layer: 20 variables};
%    \node[annot,right of=hl] {Output layer: 5 alternatives};
\end{tikzpicture}

% visualize neural net
}\label{fig:s_arch0}} 
\subfloat[F1 Deep Dense Feedforward NN (DNN)]{\resizebox{0.6\linewidth}{!}{% visualize neural net
\def\layersep{1.5cm}
\centering
\begin{tikzpicture}[shorten >=1pt, ->, draw=black!50, node distance=\layersep]
    \tikzstyle{every pin edge}=[<-, shorten <=1pt]
    \tikzstyle{neuron}=[circle,fill=black!25,minimum size=17pt,inner sep=0pt]
    \tikzstyle{input neuron}=[neuron, fill=black!50];
    \tikzstyle{output neuron}=[neuron, fill=red!50];
    \tikzstyle{hidden neuron}=[neuron, fill=blue!50];
    \tikzstyle{annot} = [text width=3cm, text centered]

    % Draw the input layer nodes
%    \foreach \name / \y in {1,...,4}
    % This is the same as writing \foreach \name / \y in {1/1,2/2,3/3,4/4}
    \node[input neuron, pin=left:X0] (X0-0) at (0 cm, 0 cm) {};
    \filldraw [black!50] (0, -0.5cm) circle (2pt);
    \filldraw [black!50] (0, -1.0cm) circle (2pt);
    \node[input neuron] (X0-1) at (0 cm,-1.5cm) {};

    \node[input neuron, pin=left:X1] (X1-0) at (0 cm, -2.5cm) {};
    \filldraw [black!50] (0, -3.0cm) circle (2pt);
    \filldraw [black!50] (0, -3.5cm) circle (2pt);
    \node[input neuron] (X1-1) at (0 cm,-4.0cm) {};

    \node[input neuron, pin=left:Z] (Z-0) at (0 cm, -5.0cm) {};
    \filldraw [black!50] (0, -5.5cm) circle (2pt);
    \filldraw [black!50] (0, -6.0cm) circle (2pt);
    \node[input neuron] (Z-1) at (0 cm,-6.5cm) {};

    % Draw the hidden layer nodes
    \foreach \name / \y in {1,...,7}
        \path[yshift=2.0cm]
            node[hidden neuron] (H0-\name) at (\layersep, -\y cm) {};
    \filldraw [blue!50] (\layersep, -5.5cm) circle (2pt);
    \filldraw [blue!50] (\layersep, -6.0cm) circle (2pt);
    \filldraw [blue!50] (\layersep, -6.5cm) circle (2pt);
    \node[hidden neuron] (H0-8) at (\layersep,-7.0cm) {};

    \foreach \name / \y in {1,...,7}
        \path[yshift=2.0cm]
            node[hidden neuron] (H1-\name) at (2*\layersep, -\y cm) {};
    \filldraw [blue!50] (2*\layersep, -5.5cm) circle (2pt);
    \filldraw [blue!50] (2*\layersep, -6.0cm) circle (2pt);
    \filldraw [blue!50] (2*\layersep, -6.5cm) circle (2pt);
    \node[hidden neuron] (H1-8) at (2*\layersep,-7.0cm) {};

    \foreach \name / \y in {1,...,7}
        \path[yshift=2.0cm]
            node[hidden neuron] (H2-\name) at (3*\layersep, -\y cm) {};
    \filldraw [blue!50] (3*\layersep, -5.5cm) circle (2pt);
    \filldraw [blue!50] (3*\layersep, -6.0cm) circle (2pt);
    \filldraw [blue!50] (3*\layersep, -6.5cm) circle (2pt);
    \node[hidden neuron] (H2-8) at (3*\layersep,-7.0cm) {};

    \foreach \name / \y in {1,...,7}
        \path[yshift=2.0cm]
            node[hidden neuron] (H3-\name) at (4*\layersep, -\y cm) {};
    \filldraw [blue!50] (4*\layersep, -5.5cm) circle (2pt);
    \filldraw [blue!50] (4*\layersep, -6.0cm) circle (2pt);
    \filldraw [blue!50] (4*\layersep, -6.5cm) circle (2pt);
    \node[hidden neuron] (H3-8) at (4*\layersep,-7.0cm) {};

    \foreach \name / \y in {1,...,7}
        \path[yshift=2.0cm]
            node[hidden neuron] (H4-\name) at (5*\layersep, -\y cm) {};
    \filldraw [blue!50] (5*\layersep, -5.5cm) circle (2pt);
    \filldraw [blue!50] (5*\layersep, -6.0cm) circle (2pt);
    \filldraw [blue!50] (5*\layersep, -6.5cm) circle (2pt);
    \node[hidden neuron] (H4-8) at (5*\layersep,-7.0cm) {};

    \foreach \name / \y in {1,...,7}
        \path[yshift=2.0cm]
            node[hidden neuron] (H5-\name) at (6*\layersep, -\y cm) {};
    \filldraw [blue!50] (6*\layersep, -5.5cm) circle (2pt);
    \filldraw [blue!50] (6*\layersep, -6.0cm) circle (2pt);
    \filldraw [blue!50] (6*\layersep, -6.5cm) circle (2pt);
    \node[hidden neuron] (H5-8) at (6*\layersep,-7.0cm) {};

    \foreach \name / \y in {1,...,7}
        \path[yshift=2.0cm]
            node[hidden neuron] (H6-\name) at (7*\layersep, -\y cm) {};
    \filldraw [blue!50] (7*\layersep, -5.5cm) circle (2pt);
    \filldraw [blue!50] (7*\layersep, -6.0cm) circle (2pt);
    \filldraw [blue!50] (7*\layersep, -6.5cm) circle (2pt);
    \node[hidden neuron] (H6-8) at (7*\layersep,-7.0cm) {};

    % Draw the output layer node
    \foreach \name / \y in {0,...,1}
        \node[output neuron, pin={[pin edge={->}]right:Y \y}] (Y-\name) at (8*\layersep, -\y cm - 2.5cm) {};

    % Connect every node in the hidden layer with the output layer
    \foreach \source in {0,...,1}
        \foreach \dest in {1,...,8}
            \path (X0-\source) edge (H0-\dest);

    \foreach \source in {0,...,1}
        \foreach \dest in {1,...,8}
            \path (X1-\source) edge (H0-\dest);

    \foreach \source in {0,...,1}
        \foreach \dest in {1,...,8}
            \path (Z-\source) edge (H0-\dest);

    \foreach \source in {1,...,8}
        \foreach \dest in {1,...,8}
            \path (H0-\source) edge (H1-\dest);

    \foreach \source in {1,...,8}
        \foreach \dest in {1,...,8}
            \path (H1-\source) edge (H2-\dest);

    \foreach \source in {1,...,8}
        \foreach \dest in {1,...,8}
            \path (H2-\source) edge (H3-\dest);

    \foreach \source in {1,...,8}
        \foreach \dest in {1,...,8}
            \path (H3-\source) edge (H4-\dest);

    \foreach \source in {1,...,8}
        \foreach \dest in {1,...,8}
            \path (H4-\source) edge (H5-\dest);            

    \foreach \source in {1,...,8}
        \foreach \dest in {1,...,8}
            \path (H5-\source) edge (H6-\dest);            

    \foreach \source in {1,...,8}
        \foreach \dest in {0,...,1}
            \path (H6-\source) edge (Y-\dest);

%    \foreach \source in {1,...,5}
%        \path (H-\source) edge (O);

    % Annotate the layers
%    \node[annot,left of=hl] {Input layer: 20 variables};
%    \node[annot,right of=hl] {Output layer: 5 alternatives};
\end{tikzpicture}}\label{fig:s_arch1}} \\ 
\caption{Two Architectures of BNL and DNN; first graph represents BNL with linear specification, second graph is DNN. Visually, DNN is an extension of BNL, as is its function class. The red neurons in both graphs visualize utility values, and the blue neurons in DNN can be seen as the process of specifying utility.}
\label{fig:two_arch}
\end{figure}
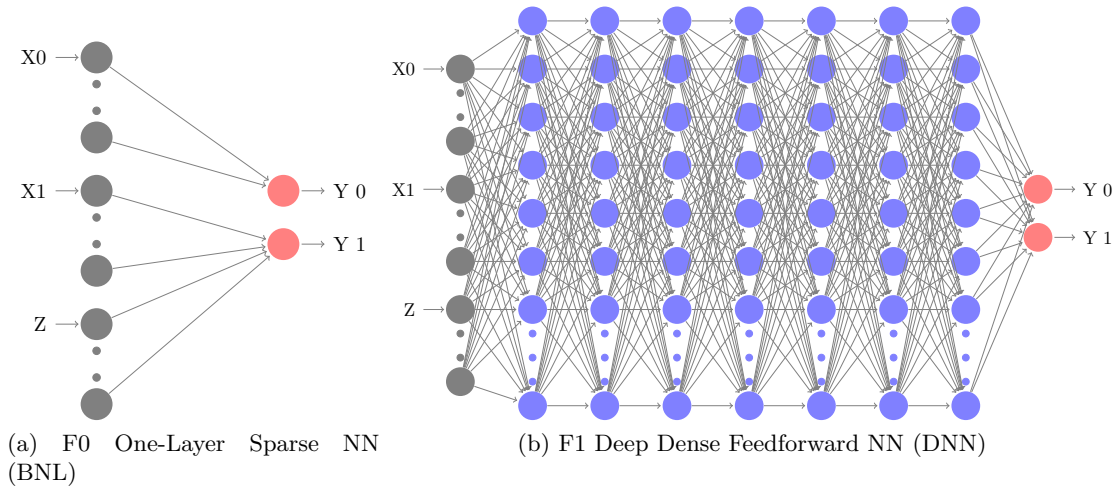

\noindent
Since BNL is one subset of DNN (${\F}_0 \subset {\F}_1$) (shown in Figure \ref{fig:two_arch}), the approximation error of DNN is always smaller than BNL \cite{Vapnik1999}. Intuitively, the best model ($f^*_{{\F}_0}$) in ${\F}_0$ is also in ${\F}_1$, so it is generally true that $f^*_{{\F}_1}$ can approximate $f^*$ better than $f^*_{{\F}_0}$. Formally,

\begin{proposition} \label{prop:approx_error_prediction}
The approximation error of the prediction loss in DNN is always smaller than that in BNL
\begin{flalign}
& {\E}_S[L_{0/1}(f^*_{\F_1}) - L_{0/1}(f^*)] \leq {\E}_S[L_{0/1}(f^*_{\F_0}) - L_{0/1}(f^*)]
\end{flalign}
Similarly, the approximation error of the interpretation loss in DNN is also smaller than that in BNL:
\begin{flalign}
& {\E}_S[L_e(s^*_{\F_1}) - L_e(s^*)] \leq {\E}_S[L_e(s^*_{\F_0}) - L_e(s^*)]
\end{flalign}
\end{proposition}

\noindent
While these results are not difficult to see, they can be understood from various mathematical perspectives. The first perspective is the \textit{universal approximator theorem} of DNN, developed in the 1990s. The studies suggest that even a shallow neural network (SNN) is asymptotically a universal approximator when the width becomes infinite \cite{Cybenko1989,Hornik1989,Hornik1991}. Recently, this asymptotic perspective leads to a more non-asymptotic question, asking why depth is necessary for SNN to be powerful enough for practical use cases. Research has demonstrated that DNN can approximate functions with an exponentially smaller number of neurons than SNN in many settings \cite{Cohen2016,Rolnick2017,Poggio2017}. This perspective is quite relevant to our focus, since BNL is one type of SNN \cite{Bentz2000}. The choice between DNN and BNL can equivalently be framed as the choice between DNN and SNN.

In addition to these mathematical perspectives, we highlight the economic perspective that describes the similarity between BNL and DNN, as well as their difference between \textit{automatic} and \textit{handcrafted} utility specification. BNL and DNN are categorically similar since both involve the process of specifying and comparing utility values. The notion of utility specification and comparison in choice modeling setting is widely known \cite{Train2009,Ben_Akiva1985}, but they can also be applied to DNN. In fact, the last layer of DNN could be named as utilities and the layers before the last can be seen as utility specifications. However, their key difference is that traditional BNL approaches use handcrafted utility specification based on experts' prior knowledge, while DNN automatically learns utility specification based on a complex model assumption. Therefore, while our study discusses only the linear utility specification for BNL, BNL and DNN broadly embody two paradigms of utility specification. Any form of handcrafted features, such as incorporating quadratic or log terms, can always be used as inputs into DNN, enabling additional automatic learning power. Automatic feature learning is nearly inevitable in many tasks such as face recognition, in which handcrafting features of human faces seem nearly impossible \cite{Mullainathan2017}. Studies in the ML community typically praise the power of this automatic feature learning, although it is still a heated debate whether researchers should rely on only the automatic feature learning or a mixture of automatic and handcrafted feature learning \cite{LeCun2015,Bengio2013,Qianli2018}. However, the bottom line is that a pure handcrafted utility specification will not be able to maximize the predictive and interpretable power of the data, and using or at least augmenting the power of the automatic feature learning in DNN could greatly add to the future of choice modeling practice.

\subsection{Estimation Error}
\noindent
The more challenging question is about the estimation error of DNN, particularly because the smaller approximation error is always associated with larger estimation errors. Specifically, the question is whether DNN has well-bounded estimation error, when the number of its parameters is so large. To address this question, we will present two proofs. While both rely on empirical process theory, the first uses contraction inequality, which provides a tighter upper bound than the second proof, which is based on VC dimension. The proof based on  empirical process theory shows that the estimation error of both prediction and interpretation losses in DNN can be bounded or at least controlled by $l_1$ and $l_2$ regularizations. We believe this part is critical since the empirical process theory provides a new foundation for future studies that rely on high-dimensional statistical tools used for individual choice modeling. We put only the key propositions in the following section, with detailed proofs provided in Appendix II.

\begin{definition} \label{def:rad_complexity}
Empirical Rademacher complexity is defined as 
\begin{flalign}
\hat{\R}_n({\F}|_S) = {\E}_{\epsilon} \Big[ \ \underset{f \in {\F}}{\sup} \ \Big| \frac{1}{N} \sum_{i=1}^N \epsilon_i f(x_i) \Big| \Big]
\end{flalign}
$\epsilon_i \in \{+1,-1 \}$ with probabilties $[0.5, 0.5]$; ${\F}|_S$ denotes the function class $\F$ projected to sample $S$.
\end{definition}

\begin{proposition} \label{prop:excess_error_rad_complexity}
The estimation error of $\hat{f}$ can be upper bounded by the Rademacher complexity
\begin{flalign}
{\E}_S [L(\hat f) - L(f_F^*)] \leq 2 {\E}_S \hat{\R}_n(l \circ {\F}|_S)
\label{eq:rad_upper_bound}
\end{flalign}
\end{proposition}

\noindent
Proof of Proposition \ref{prop:excess_error_rad_complexity} is available in Appendix II.A. Rademacher complexity measures the complexity of function class ${\F}$ conditioning on the dataset $S$. Proposition \ref{prop:excess_error_rad_complexity} shows that estimation error can be upper bounded by the complexity of the function class $l \circ \F$, defined as $l \circ \F = \{l \circ f(x) | f(x) \in \F \}$. Intuitively, it is harder to search for the best model $\hat{f}$, as the function class $\F$ becomes larger. It also aligns with traditional statistics, as higher VC dimension or more parameters (more complexity of function class) leads to larger estimation errors. Specifically, Proposition \ref{prop:excess_error_rad_complexity} can be used as an upper bound for the estimation errors of prediction and interpretation losses:

\begin{proposition} \label{prop:excess_error_rad_complexity_prediction}
The estimation error of the prediction loss can be (indirectly) upper bounded
\begin{flalign} \label{eq:prediction_error_bound}
{\E}_S[L_{0/1}(\hat f) - \hat{L}_{\gamma}(\hat{f})] \leq \frac{2}{\gamma} {\E}_S \hat{\R}_n({\F}|_S)
\end{flalign}
\end{proposition}

\begin{proposition} \label{prop:excess_error_rad_complexity_interpretation}
The estimation error of the interpretation loss can be upper bounded by 
\begin{flalign} \label{eq:interpretation_error_bound}
{\E}_S[L_{e}(\hat s) - L_{s}(s_F^*)] \leq 4 {\E}_S \hat{\R}_n({\F}|_S)
\end{flalign}
\end{proposition}

\noindent
Proof of Propositions \ref{prop:excess_error_rad_complexity_prediction} and \ref{prop:excess_error_rad_complexity_interpretation} is available in Appendix II.B and II.C. Proposition \ref{prop:excess_error_rad_complexity_prediction} provides an upper bound on ${\E}_S[L_{0/1}(\hat f)]$ by using $\gamma$-margin error (definition in Appendix II.B). While the left hand side is not exactly the same as ${\E}_S[L_{0/1}(\hat f) - L_{0/1}(f^*_F)]$, both $\hat{L}_{\gamma}(\hat{f})$ and $\frac{2}{\gamma} {\E}_S \hat{\R}_n({\F}|_S)$ can be computed in practice. Compared to the estimation error of the prediction loss, the interpretation part is easier, and Proposition \ref{prop:excess_error_rad_complexity_interpretation} demonstrate that the estimation error of the interpretation loss is upper bounded by Rademacher complexity up to a constant. One remaining question is how to provide an effective upper bound on Rademacher complexity of DNN. 

\begin{proposition} \label{prop:rad_complexity_dnn_contraction}
Let $H_d$ be the class of neural network with depth $D$ over the domain $\X$ ($x \in B_1^{(d_0)}$), where each parameter matrix $W_j$ has Frobenius norm at most $M_F(j)$ and its one-infinity norm at most $M(j)$, and with ReLU activation functions. Then by using contraction inequality, the Rademacher complexity of DNN (${\F}_1$) can be upper bounded by 
\begin{flalign}
\hat{\R}_n({\F}_1|_S) \lesssim O(\frac{\sqrt{\log d_0} \times \prod_{j = 1}^D 2M(j)}{\sqrt{N}})
\end{flalign}
\noindent
The tightest bound found in the literature \cite{Golowich2017} is:
\begin{flalign} \label{eq:rad_complexity_dnn_contraction}
\hat{\R}_n({\F}_1|_S) \lesssim  \frac{\sqrt{\log d_0} \times (\sqrt{2 \log D} + 1) \times \prod_{j = 1}^D M_F(j)}{\sqrt{N}}
\end{flalign}
\end{proposition}

\begin{proposition} \label{prop:rad_complexity_dnn_vc}
Rademacher complexity of DNN with $0/1$ loss can be upper bounded by VC dimension
\begin{flalign} \label{eq:rad_complexity_dnn_vc}
\hat{\R}_n(l \circ {\F}_1) \lesssim 4 \sqrt{\frac{v \log(N + 1)}{N}} \lesssim 4 \sqrt{\frac{TD \log(T) \times \log(N + 1)}{N}} 
\end{flalign}
\noindent
with $T$ denoting the total number of parameters and $D$ the depth of DNN \cite{Bartlett2017}.
\end{proposition}

\noindent
Proposition \ref{prop:rad_complexity_dnn_contraction} describes the important factors that influence the upper bound on estimation error, including the input dimension $d_0$, norm of parameters in each layer $M(j)$ or $M_F(j)$, and sample size. The result is intuitive: with larger sample size, and smaller input dimension and norm of parameters, the estimation error of DNN is more likely to be bounded. Proof of Propositions \ref{prop:rad_complexity_dnn_contraction} and \ref{prop:rad_complexity_dnn_vc} is available in Appendix II.D and II.E. 

The most important message about estimation error is revealed by the difference between Propositions \ref{prop:rad_complexity_dnn_contraction} and \ref{prop:rad_complexity_dnn_vc}: instead of computing the ratio of $v$ and $N$ as in Proposition \ref{prop:rad_complexity_dnn_vc}, researchers could compute the \textit{norms} of the coefficients in each layer to upper bound estimation error, as in Proposition \ref{prop:rad_complexity_dnn_contraction}. The total number of parameters is fixed when researchers choose one specific DNN architecture, so it is hard to control the Rademacher complexity through VC dimension. On the contrary, the norms of the weights in each layer $M(j)$ can be controlled by $l_1$ or $l_2$ regularization. Therefore, Proposition \ref{prop:rad_complexity_dnn_contraction} along with Propositions \ref{prop:excess_error_rad_complexity_prediction} and \ref{prop:excess_error_rad_complexity_interpretation} provide valid and much tighter upper bounds on estimation error than the traditional VC dimension perspective. 

The results above heavily rely on the progresses in non-asymptotic statistical learning theory and particularly the empirical process theory in the recent two decades. Readers should refer to  \cite{Bousquet2004,Von_Luxburg2011,Anthony2009,Wainwright2019,Vapnik2013} for general introductions; to \cite{Vapnik1999,Vapnik2013,Sontag1998,Bartlett2017} for the proof about Rademacher complexity bound of DNN based on VC dimension; to \cite{Golowich2017,Neyshabur2015,Bartlett2002,Bartlett2006} for the proof about Rademacher complexity bound of DNN based on contraction inequality.

\subsection{Summary}
\noindent
So far we have provided concrete mathematical formulation and theoretical discussions for the two dimensions and four quadrants that define our theoretical framework, as summarized in Table \ref{table:two_dim_theory}. Both dimensions are important from a historical view. The tradeoff between estimation and approximation error is the first order decomposition in statistical learning theory \cite{Vapnik1999,Vapnik2013,Von_Luxburg2011}. Prediction vs. interpretation marks the difference of two statistical cultures, as pointed out by Leo Breiman (2001) \cite{Breiman2001}, and is recently remarked again by Mullainathan and Spiess (2017) \cite{Mullainathan2017}. For the purpose of our study, the two dimensions can be used to bridge the classical low-dimensional DCMs and the new high-dimensional DNN models from a theoretical perspective.

\begin{table}[ht]
\centering
\resizebox{0.8\linewidth}{!}{
\begin{tabular}{ p{0.24\linewidth} | P{0.28\linewidth} | P{0.28\linewidth} }
\toprule
 & \textbf{Approximation Error} & \textbf{Estimation Error} \\
\hline
\textbf{Prediction Loss} & Approximation error of prediction loss ${\E}_S[L_{0/1}(f^*_F) - L_{0/1}(f^*)]$ & Estimation error of prediction loss ${\E}_S[L_{0/1}(\hat f) - L_{0/1}(f^*_F)]$ \\
\hline
\textbf{Interpretation Loss} & Approximation error of interpretation loss ${\E}_S[L_{e}(s^*_F) - L_{e}(s^*)]$ & Estimation error of interpretation loss ${\E}_S[L_{e}(\hat s) - L_{e}(s^*_F)]$ \\
\bottomrule
\end{tabular}
} %resize used here
\caption{Two Dimensions of the Theoretical Framework}
\label{table:two_dim_theory}
\end{table}

\section{Experiments}
\subsection{Design of Experiments}
\noindent
The experiments consist of two parts: one with three simulated datasets and one with the NHTS dataset. The experiments with simulated and real datasets are complementary in terms of their purposes. With Monte Carlo simulation, the underlying true data generating process (DGP; e.g. $s^*(x)$ or $f^*(x)$) is known, so we can compute both the approximation and estimation errors related to $s^*(x)$ and $f^*(x)$, which cannot be done in the experiment with real datasets. On the other hand, real datasets reveal the real decision making process, which has to be presumed, sometimes arbitrarily, in Monte Carlo simulations.

In both experiments, we compare one DNN architecture with fixed hyper-parameters to one BNL model with linear utility specification. The DNN architecture has $5$ layers, $100$ neurons in each layer, and ReLU activation functions. The DNN training uses standard ERM procedure, with He initialization \cite{He2015}, Adam optimization \cite{Kingma2014}, and mild regularizations. The BNL in all our simulations uses only linear specification. Again, this linear specification of BNL does not limit the generality of our discussion since any domain knowledge based utility specification could always be provided to DNN as inputs. DNN's theoretical properties do not vary much with the specific choice of parameters and hyperparameters. BNL and DNN broadly represent the difference between handcrafted and automatic utility specification, and the specific choice of BNL models and DNNs do not matter for the purpose of this study.

The experiment with Monte Carlo simulation consists of three scenarios, representing three typical cases researchers face in reality. The three scenarios are differentiated by the ``location'' of the true DGP with respect to ${\F}_0$ and ${\F}_1$: (1) $f^* \in F_0$ and $f^* \in F_1$; (2) $f^* \not\in F_0$ and $f^* \in F_1$; (3) $f^* \not\in F_0$ and $f^* \not\in F_1$. Scenario 1 represents the case in which a simple BNL is the true DGP, which belongs to both model classes of BNL and DNN, so the approximation errors of both BNL and DNN are zero. Secnario 2 represents the case in which the true DGP is more complicated than BNL, so the approximation error of BNL is larger than zero while that of DNN is still zero. Scenario 2 commonly happens when information is complete while the function used in model training is misspecified in choice modeling. Scenario 3 represents the case in which both BNL and DNN have strictly positive approximation errors, which happens when important variables are omitted, traditionally called omitted variable bias. In terms of function relationship between ${\F}_0$, ${\F}_1$ and $f^*$, the three scenarios are exhaustive. Our simulation also varies sample size and number of input variables to demonstrate how estimation error changes, based on our theory about estimation error of DNN (Proposition \ref{prop:rad_complexity_dnn_contraction}). Sample size in the Monte Carlo simulations ranges from $100$, the smallest possible one in a survey, to $1$ million, the largest number observed in existing transportation questionnaire-based or observational surveys. The number of input variables is either $20$ or $50$, typical in choice analysis. For each experiment, we analyze the four quadrants, estimation and approximation errors of prediction and interpretation losses, mapping back to our theoretical framework in Table \ref{table:two_dim_theory}. More details of the simulation are attached in Appendix III.

The experiment with the NHTS dataset analyzes travel mode choices and trip purpose choices, two prevalent travel behaviors analyzed in the past studies \cite{Ortuzar2011,Zegras2010,Cervero1997_3d}, with varying sample size from $100$ to $500,000$. The NHTS dataset is chosen since it covers the whole United States and it is one of the only datasets that has a sample size on the order of magnitude of $1$ million. Because of the absence of a true data-generating process, the decomposition of estimation and approximation errors is impossible for the experiment with the real datasets, but we discuss both the prediction and interpretation of DNN-based choice models.

\subsection{Three experiments with Simulated Datasets}
\subsubsection{Scenario 1}
\begin{figure}[t!]
\centering
\subfloat[Prediction Loss (20 Var)]{\includegraphics[width=0.25\linewidth]{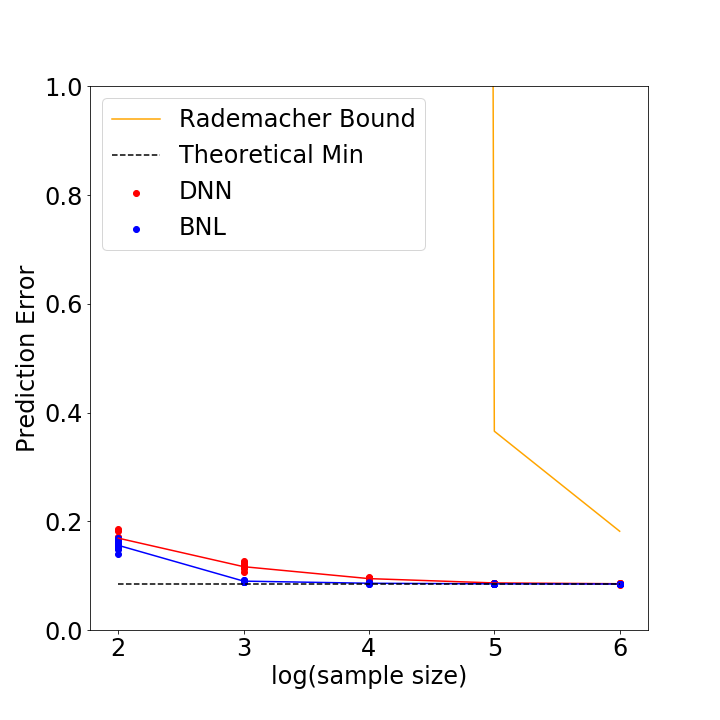}\label{sfig:s1_acc_20_var}}
\subfloat[Prediction Loss (50 Var)]{\includegraphics[width=0.25\linewidth]{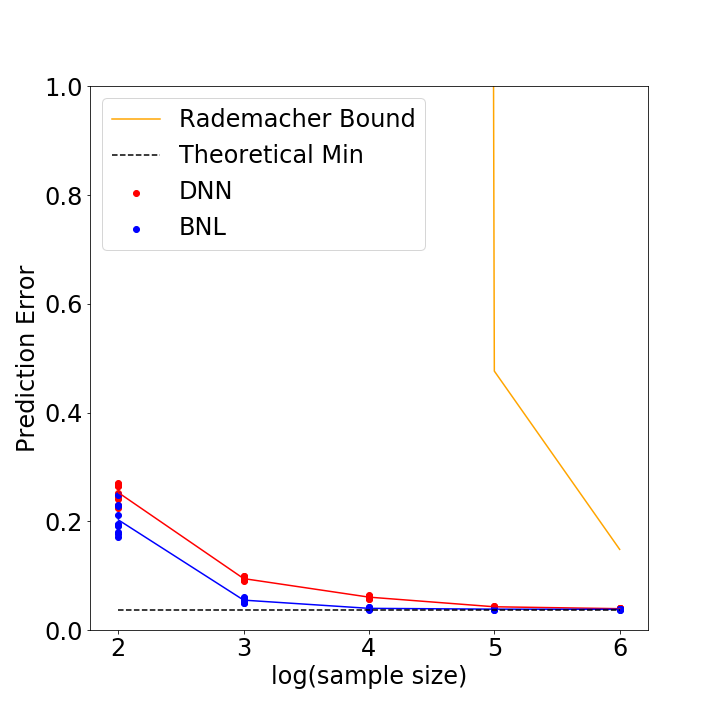}\label{sfig:s1_acc_50_var}} 
\subfloat[Interpretation Loss (20 Var)]{\includegraphics[width=0.25\linewidth]{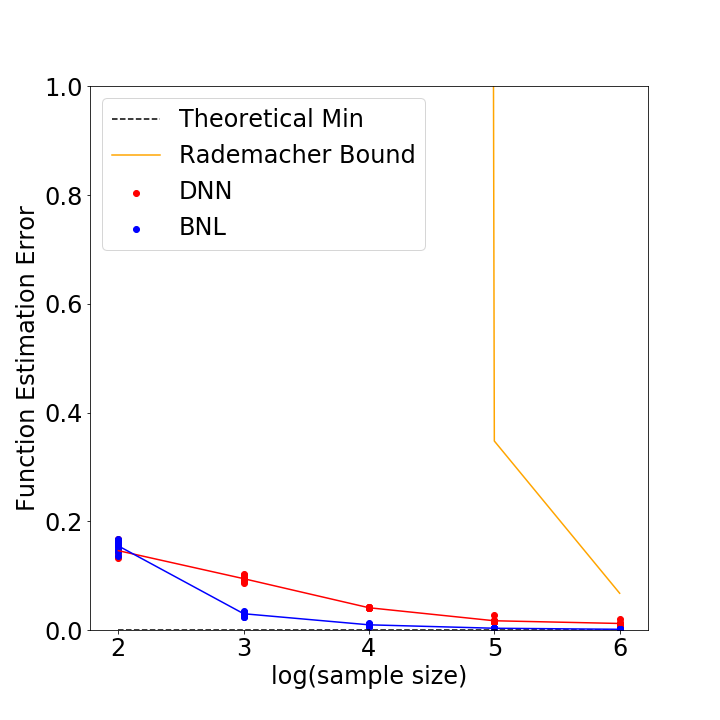}\label{sfig:s1_est_20_var}}
\subfloat[Interpretation Loss (50 Var)]{\includegraphics[width=0.25\linewidth]{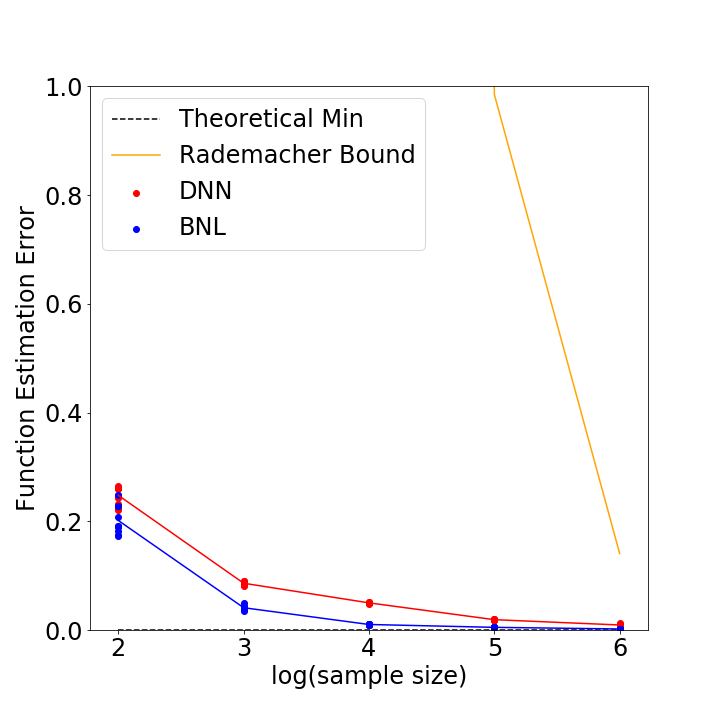}\label{sfig:s1_est_50_var}} \\
\subfloat[Choice Probability Curves (20 Var); Sample size = 100, 1000, 10000, 100000, 1000000]{\includegraphics[width=\linewidth]{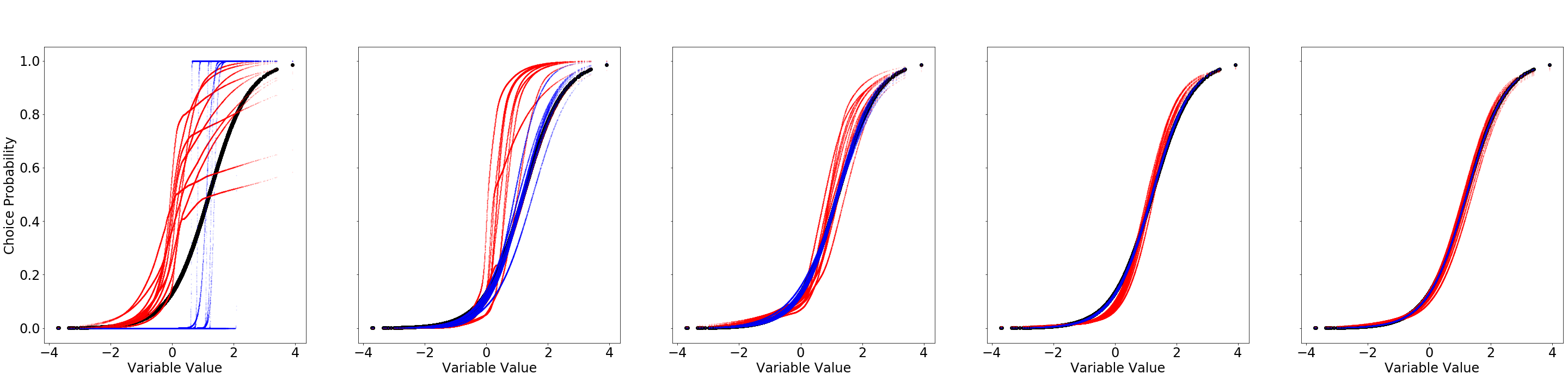}\label{sfig:s1_chprob_20_var}} \\
\caption{Scenario 1. Upper row: comparison of DNN and BNL for prediction and interpretation losses; lower row: visualizing how choice probabilities change with inputs; red curves: DNN, blue curves: BNL, black curves: true models. The figures in the upper row map to the theoretical framework in Table \ref{table:two_dim_theory}: the difference between red and black curves is the prediction/interpretation losses of DNN, which equal to only their estimation errors in this scenario since the approximation errors are zero.}
\label{fig:sce1}
\end{figure}

\noindent
In scenario 1, $s^*(x) = \sigma(\langle w,x \rangle)$, in which $\sigma$ is the Sigmoid function, $w$ is randomly generated variables taking $\{-1, +1\}$ values with equal probabilities, and $x$ is generated as multivariate Gaussian distribution. In Figure \ref{fig:sce1}, the upper row shows the prediction loss of simulations (Figures \ref{sfig:s1_acc_20_var} and \ref{sfig:s1_acc_50_var}) and the interpretation loss (Figures \ref{sfig:s1_est_20_var} and \ref{sfig:s1_est_50_var}) with $20$ and $50$ input variables. In each subfigure, y-axis represents the values of prediction or interpretation losses; x-axis represents sample size; each dot is a training result with the red ones representing DNN and blue ones representing BNL; red and blue curves are the average values of the losses conditioning on a sample size. The black dashed line represents the minimum possible loss, which measures the amount of randomness in each DGP. In scenario 1, the gap between the red curve and the dashed black line is the estimation error, since it is exactly ${\E}_S [L(\hat{f}) - L(f_F^*)]$ \footnote{In scenario 1, $f_F^*$ is the same as $f^*$. Hence $L(f_F^*)$ is represented by the black dash line}. The yellow curves represent the theoretical upper bound on estimation error, based on Proposition \ref{prop:rad_complexity_dnn_contraction}. The lower row of Figure \ref{fig:sce1} shows the relationship between choice probabilities and an input variable with varying sample sizes from $100$ to $1,000,000$. In each subfigure, the black curve represents the true $s^*(x)$; each red curve represents the estimated function $\hat{s}(x)$ from DNN, and each blue one represents that from BNL. 

The estimation error of both prediction and interpretation losses in both DNN and BNL converges to zero as sample size increases, and the convergence of DNN's estimation error is only slightly slower than that of BNL, as shown in Figures from \ref{sfig:s1_acc_20_var} to \ref{sfig:s1_est_50_var}. It is not surprising that estimation errors always decrease as sample sizes increase since Equations \ref{eq:prediction_error_bound} and \ref{eq:interpretation_error_bound} imply that larger sample size leads to smaller out-of-sample prediction and interpretation losses. What is surprising is that the convergence of DNN is only \textit{marginally} slower than BNL, particularly as examined from the classical statistical perspective since the number of parameters in the DNN is about $2,000$ times more than the parsimonious BNL model. More precisely, the VC dimension of our DNN architecture $v = 50,000 \times 5 \times \log(50,000) \simeq 3 Million$ (Equation \ref{eq:rad_complexity_dnn_vc}), which is larger than any sample size we use and far-off from the classical asymptotic data regime. On the contrary, the theoretical upper bound based on contraction inequality (Propositions \ref{prop:rad_complexity_dnn_contraction} and \ref{prop:rad_complexity_dnn_vc}) is represented by the yellow curve, which is much tighter than that based on the VC dimension, although it is still quite loose compared to the simulation results. Therefore, the simulation results resonate with our theoretical discussion that number of parameters in DNN should not be used to measure its estimation error bound. Empirically, DNN and BNL need roughly the same amount of data for accurate interpretation and prediction. With $20$ or $50$ variables, at least about $10^4$ samples are needed so that the prediction and interpretation losses of DNN become close to the theoretical minimum. While this $10^4$ sample size is slightly larger than the sample sizes commonly obtained by questionnaire surveys, it is not unattainable; for instance, NHTS dataset has about $700,000$ observations, which is much larger than $10^4$. 

To interpret DNN results, we visualize the relationship between $\hat{s}(x)$ and one input variable $x_j$, as shown in Figure \ref{sfig:s1_chprob_20_var}. This method of visualizing sensitivity of $\hat{s}(x)$ with respect to $x_j$ has been used for interpreting DNN results in several studies \cite{Rao1998,Bentz2000,Montavon2018}. Again, the estimated $\hat{s}(x)$ from DNN converges very quickly towards the true $s(x)$, and it captures the S-shaped choice probability curve and the linear utility specification, even when it is not \textit{a priori} specified as linear. Overall, when researchers are very confident that prior expert knowledge has captured \textit{every} piece of information, the BNL with handcrafted features perform better in prediction and interpretation, although DNN is only slightly worse.

\subsubsection{Scenario 2}
\begin{figure}[t!]
\centering
\subfloat[Prediction Loss (20 Var)]{\includegraphics[width=0.25\linewidth]{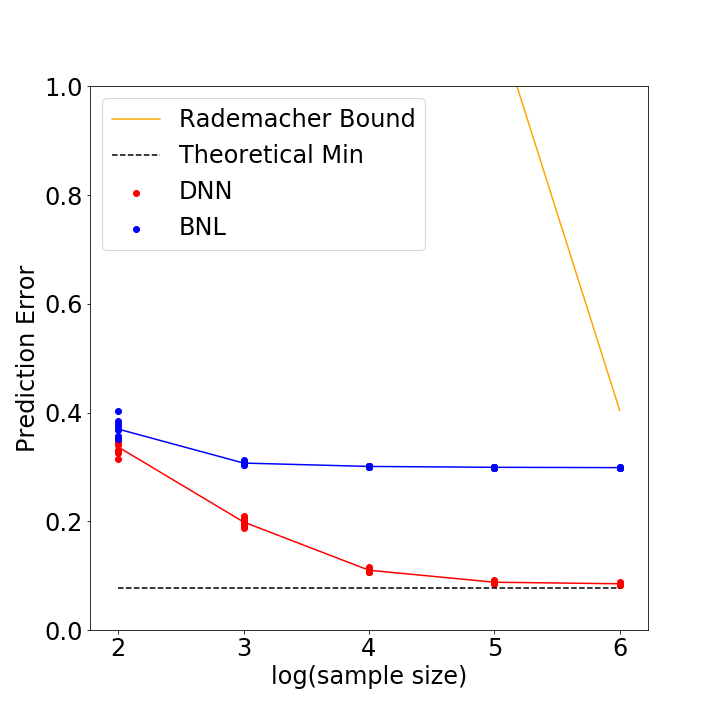}\label{sfig:s2_acc_20_var}}
\subfloat[Prediction Loss (50 Var)]{\includegraphics[width=0.25\linewidth]{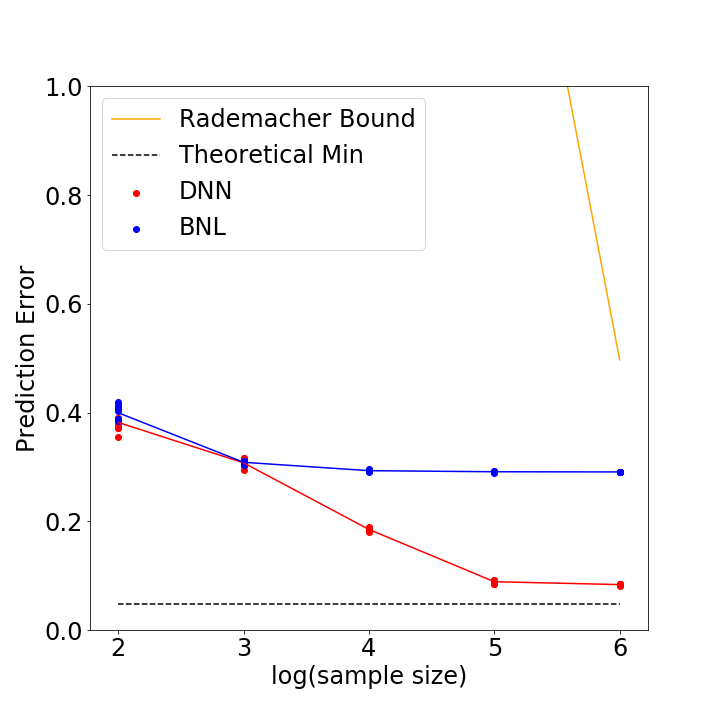}\label{sfig:s2_acc_50_var}}
\subfloat[Interpretation Loss (20 Var)]{\includegraphics[width=0.25\linewidth]{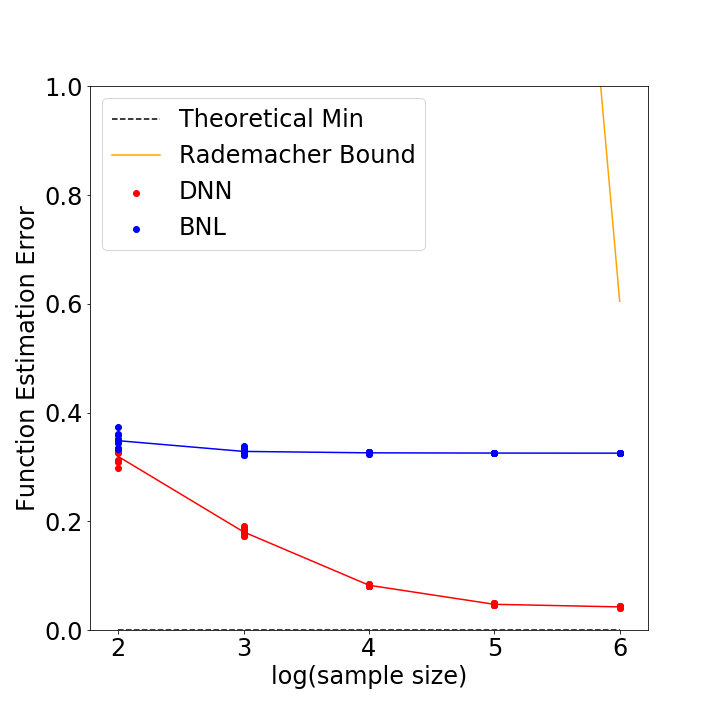}\label{sfig:s2_est_20_var}}
\subfloat[Interpretation Loss (50 Var)]{\includegraphics[width=0.25\linewidth]{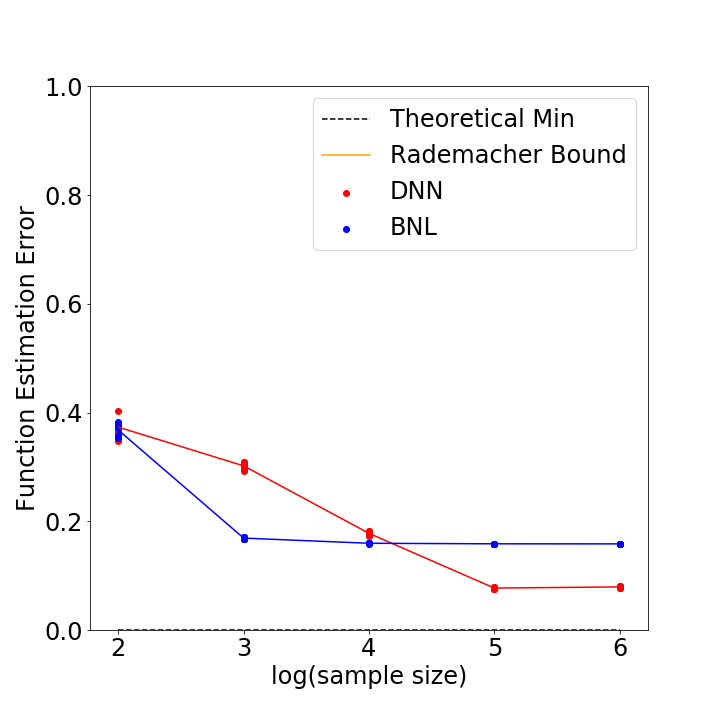}\label{sfig:s2_est_50_var}} \\
\subfloat[Choice Probability Curves (20 Var); sample size = 100, 1000, 10000, 100000, 1000000]{\includegraphics[width=\linewidth]{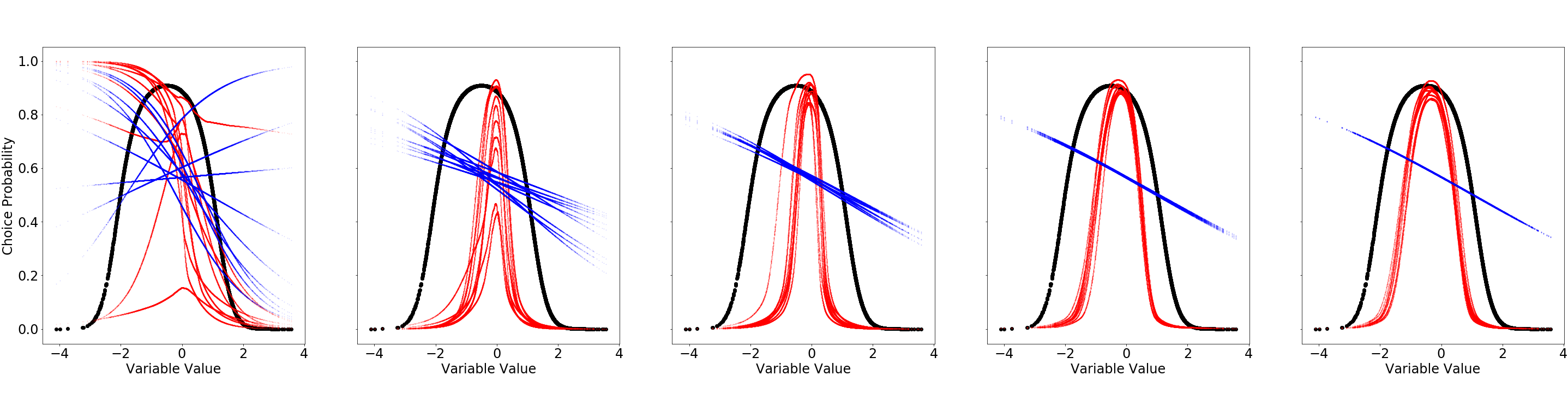}\label{sfig:s2_chprob_20_var}} \\
\caption{Scenario 2. Upper row: comparison of DNN and BNL for prediction and interpretation losses; lower row: visualizing how choice probabilities change with inputs; red curves: DNN, blue curves: BNL, black curves: true models. The figures in upper row map to the theoretical framework in Table \ref{table:two_dim_theory}. Different from Scenario 1, BNL has approximation errors since the blue curves cannot converge to the theoretical minimum values, whereas DNN has no approximation errors.}
\label{fig:sce2}
\end{figure}

\noindent
A more realistic case is the scenario in which researchers cannot correctly specify the utility function. In scenario 2, $s^{*}(x) = \sigma(w'\phi(x))$, in which $\phi(x)$ takes the quadratic transformation: $\phi([x_1, x_2, ..., x_d]) = [x_1, x_2, ... x_d, x^2_1, x^2_2, ... x_d^2 ])$. Then BNL ${\F}_0$ has the misspecification error, while DNN ${\F}_1$ does not. The results are visualized in Figure \ref{fig:sce2}, and the formats of Figure \ref{fig:sce2} is exactly the same as Figure \ref{fig:sce1}.

In scenario 2, DNN dominates BNL in terms of both prediction and interpretation losses, even at a relatively small sample size, as shown in Figures from \ref{sfig:s2_acc_20_var} to \ref{sfig:s2_est_50_var}. The key reason of DNN's dominance is its zero approximation error, in contrast to the large approximation error of BNL, measured by the gap between the theoretical minimum and the loss value that the blue curve converges to. Sample size is still a critical factor for both BNL and DNN, although it differs in terms of the critical magnitude for each. BNL converges to its optimum value ($f^*_{{\F}_0}$) with only about $10^3$ observations, while DNN converges to its optimum ($f^*_{{\F}_1} = f^*$) when sample size reaches about $10^5$ or $10^6$. This result is very consistent with our theoretical discussion. BNL aligns with classical statistics, and as $v/N$ is small, its estimation error is small. This result also implies that low dimensional statistical tools such as BNL cannot unleash the predictive power of a dataset with a large sample size. Only very complicated models such as DNN can fully unleash the predictive and interpretative power of a large sample. 

Figure \ref{sfig:s2_chprob_20_var} visualizes the relationship between $\hat{s}(x)$ and an input variable $x$ with varying sample sizes. With function misspecification, it is impossible for BNL to recover the true pattern even if the sample size becomes very large, while DNN with the power of automatic utility specification can gradually learn the underlying true utility specification, even without prior domain knowledge. Consistent with Figures \ref{sfig:s2_est_20_var} and \ref{sfig:s2_est_50_var}, DNN needs the sample size at the scale of about $10^5$ and $10^6$ to recover the true pattern of choice probability functions. Due to the misspecification and its corresponding approximation error in BNL, it is possible for DNN to outperform BNL in terms of both prediction and interpretation even when sample size is very small.

\subsubsection{Scenario 3}
\begin{figure}[t!]
\centering
\subfloat[Prediction Loss (20 Var)]{\includegraphics[width=0.25\linewidth]{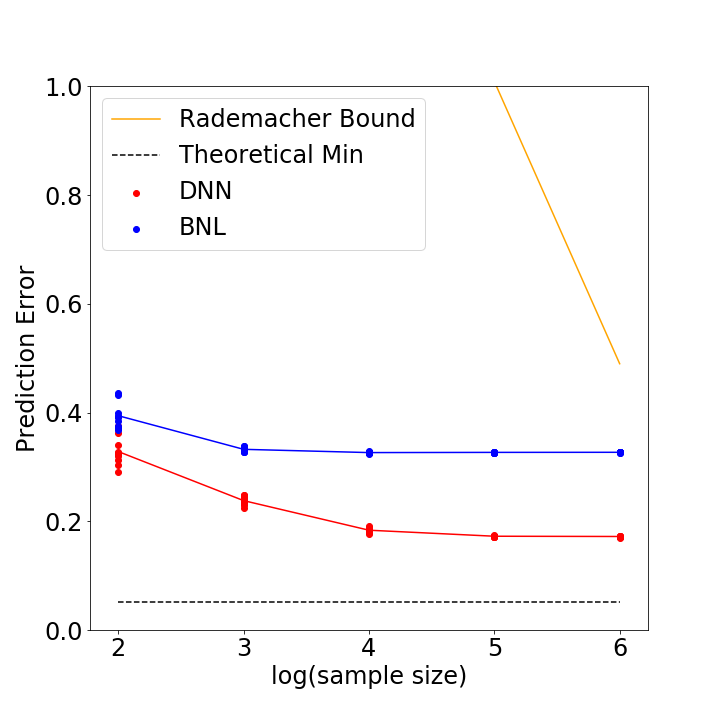}\label{sfig:s3_acc_20_var}}
\subfloat[Prediction Loss (50 Var)]{\includegraphics[width=0.25\linewidth]{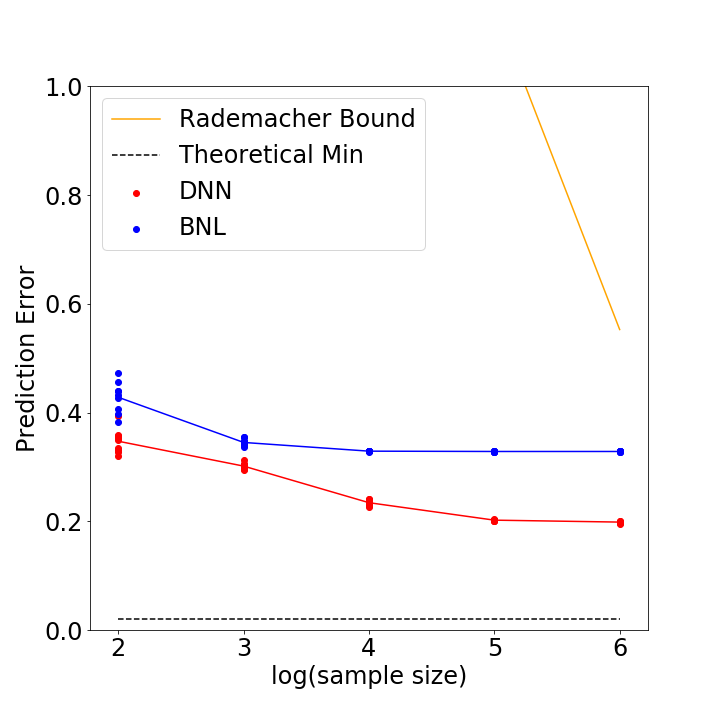}\label{sfig:s3_acc_50_var}}
\subfloat[Interpretation Loss (20 Var)]{\includegraphics[width=0.25\linewidth]{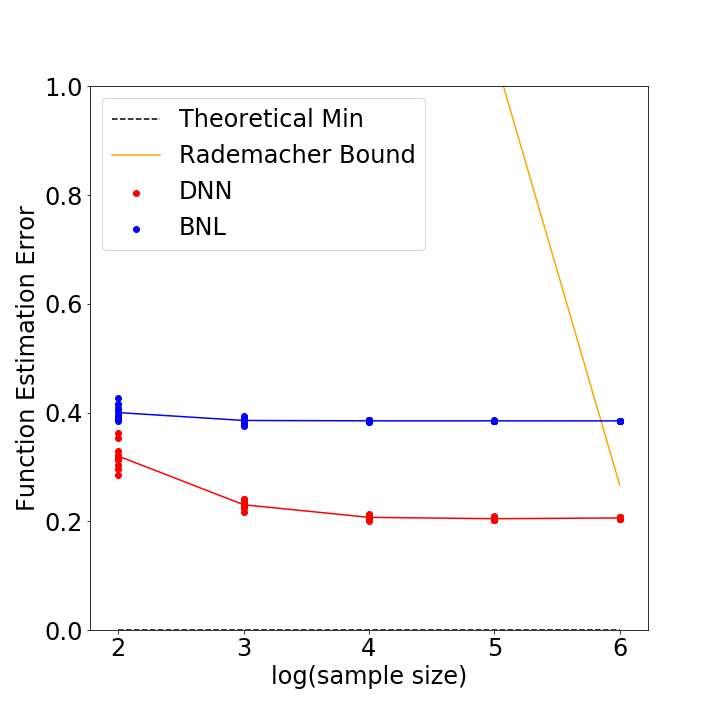}\label{sfig:s3_est_20_var}}
\subfloat[Interpretation Loss (50 Var)]{\includegraphics[width=0.25\linewidth]{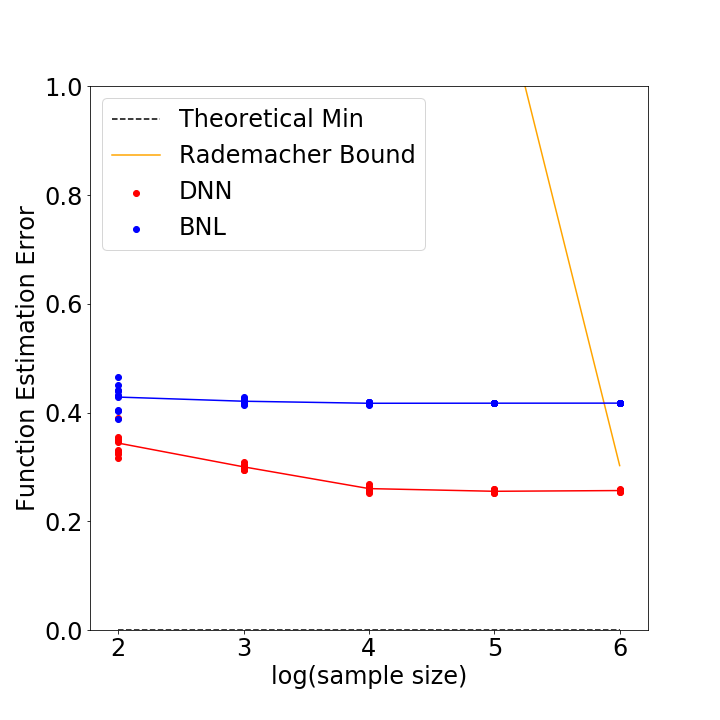}\label{sfig:s3_est_50_var}} \\
\subfloat[Choice Probability Curves (20 Var); sample size = 100, 1000, 10000, 100000, 1000000]{\includegraphics[width=\linewidth]{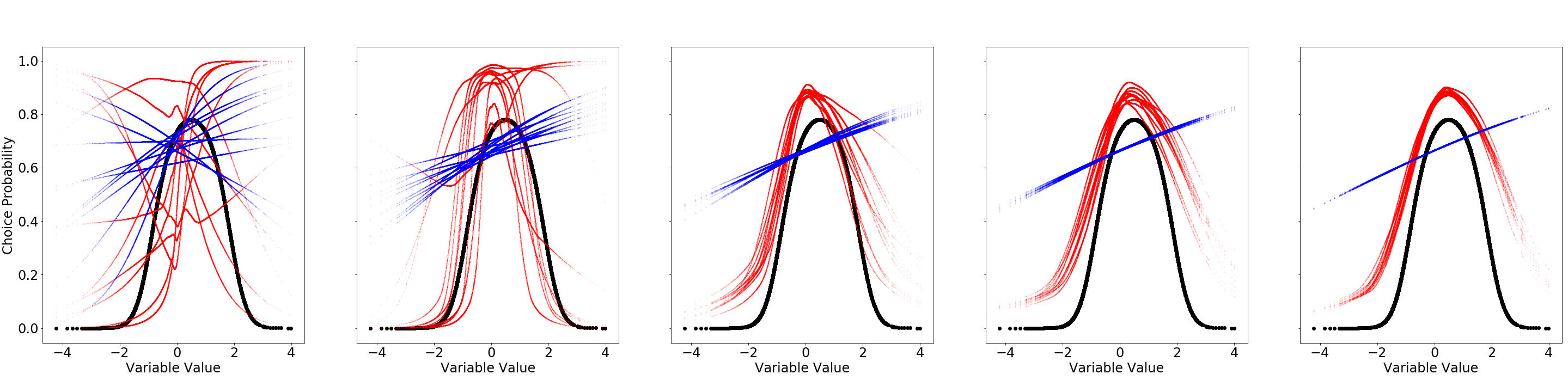}\label{sfig:s3_chprob_20_var}} \\
\caption{Scenario 3. Upper row: comparison of DNN and BNL for prediction and interpretation losses; lower row: visualizing how choice probabilities change with inputs; red curves: DNN, blue curves: BNL, black curves: true models. Different from scenario 1 and 2, both BNL and DNN have approximation errors since the red curves cannot converge to the theoretical minimum values.}
\label{fig:sce3}
\end{figure}

\noindent
A even more realistic case is the scenario in which researchers can neither collect the full information nor correctly specify the utility function ($f^* \not\in F_0$ and $f^* \not\in F_1$). In scenario 3, $s^*(x) = \sigma(w'\phi(x))$, where $\phi(x) = [1, x_1, x_2, ..., x_{d}, {x_1}^2, {x_2}^2, ... {x_d}^2, x_1x_2, ... x_{d-1}x_d]$, which includes both quadratic transformation and interaction terms. To make $f^* \not\in F_1$, we randomly drop $5$ variables out of $20$ and $20$ variables out of $50$ in training, so that even $f_{{\F}_1}^*$ cannot approximate $f^*$ well. Results are visualized in Figure \ref{fig:sce3}, with the same format as previous ones.

As shown in Figure \ref{fig:sce3}, the results are very similar to that in scenario 2, with only one critical difference that DNN also has the approximation error here. The approximation error of DNN is measured by the difference between the theoretical minimum and the values of the prediction and interpretation losses that DNNs converge to: the red curves no longer converge to theoretical minimum due to the existence of approximation errors, as shown in Figures \ref{sfig:s3_acc_20_var}-\ref{sfig:s3_est_50_var}. It is also an important message that DNN, although frequently referred to as a universal approximator, still suffers from the threat such as omitting variables. Without the completeness of information, it is unlikely for DNN to approximate the underlying $s^*(x)$ well. However, Figure \ref{sfig:s3_chprob_20_var} suggests that DNN could still well capture the choice probability function with respect to observed variables, even with omitted variables. The red curves (DNN) could approximate the true bell-shaped choice probability functions when the sample size reaches $10^4$ or $10^5$. 

Overall, this scenario demonstrates that DNN cannot solve all the problems. The ``universal approximator'' statement only applies to the functional forms of observed information, therefore only holds when all the information is observed in the model. However, even with omitted information, DNN still performs better than BNL in terms of both prediction and interpretation, owing to its power of stretching the observed information for the unobserved ones and the power of automatically learning utility specification.

\subsection{Experiment with NHTS Dataset}
\begin{figure}[t!]
\centering
\subfloat[Mode Choice Prediction (20 Variables)]{\includegraphics[width=0.25\linewidth]{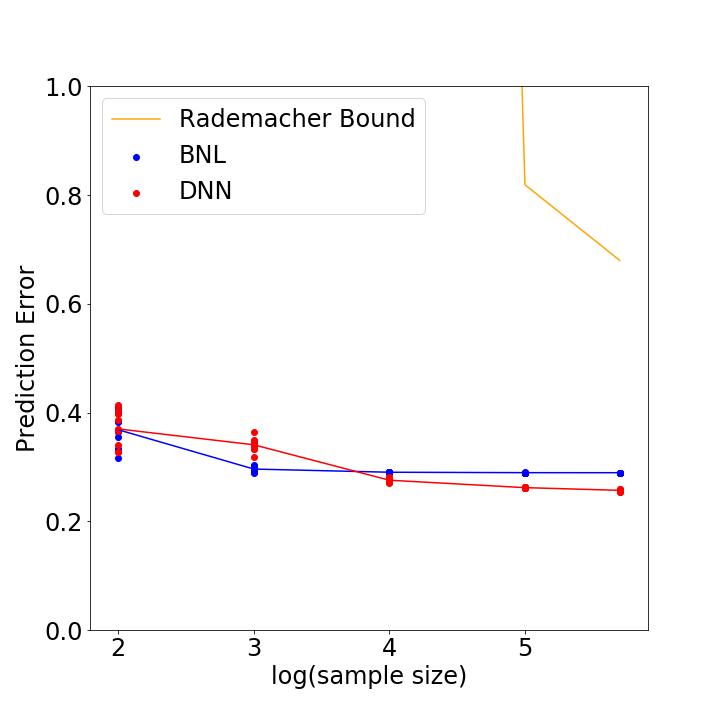}\label{sfig:nhts_mode_20_var}}
\subfloat[Mode Choice Prediction (50 Variables)]{\includegraphics[width=0.25\linewidth]{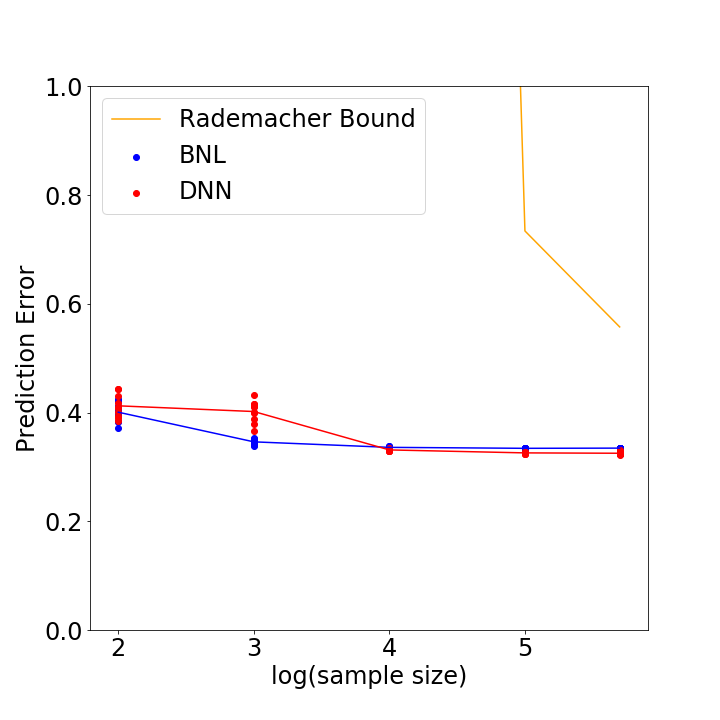}\label{sfig:nhts_mode_50_var}}
\subfloat[Trip Purpose Prediction (20 Variables)]{\includegraphics[width=0.25\linewidth]{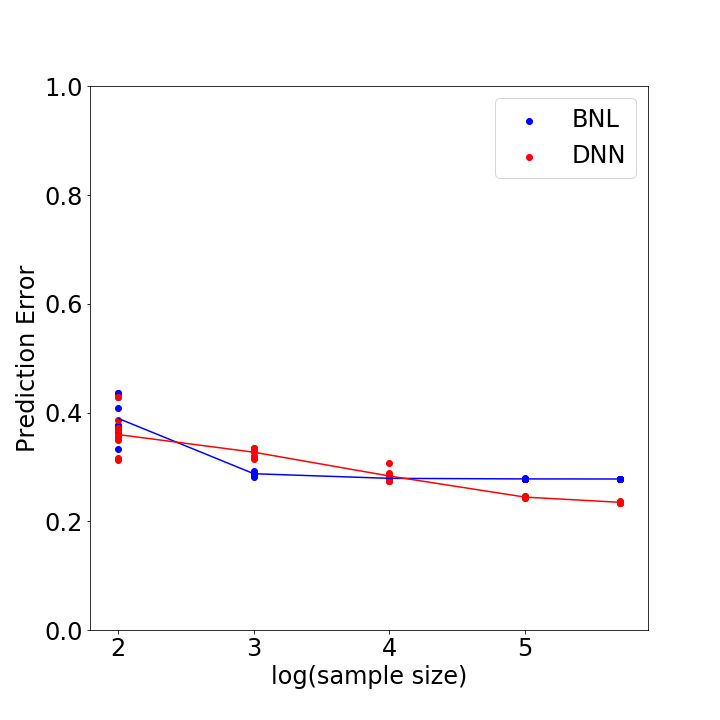}\label{sfig:nhts_purp_20_var}}
\subfloat[Trip Purpose Prediction (50 Variables)]{\includegraphics[width=0.25\linewidth]{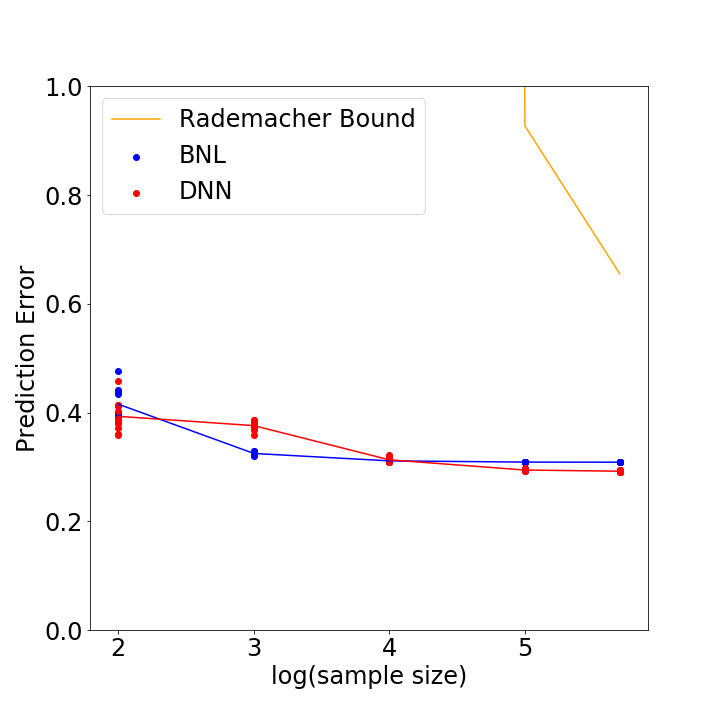}\label{sfig:nhts_purp_50_var}} \\
\subfloat[Choice Probabilility Change w.r.t. Trip Distance (From Left to Right: Sample Size 100, 1000, 10000, 100000, 500000)]{\includegraphics[width=\linewidth]{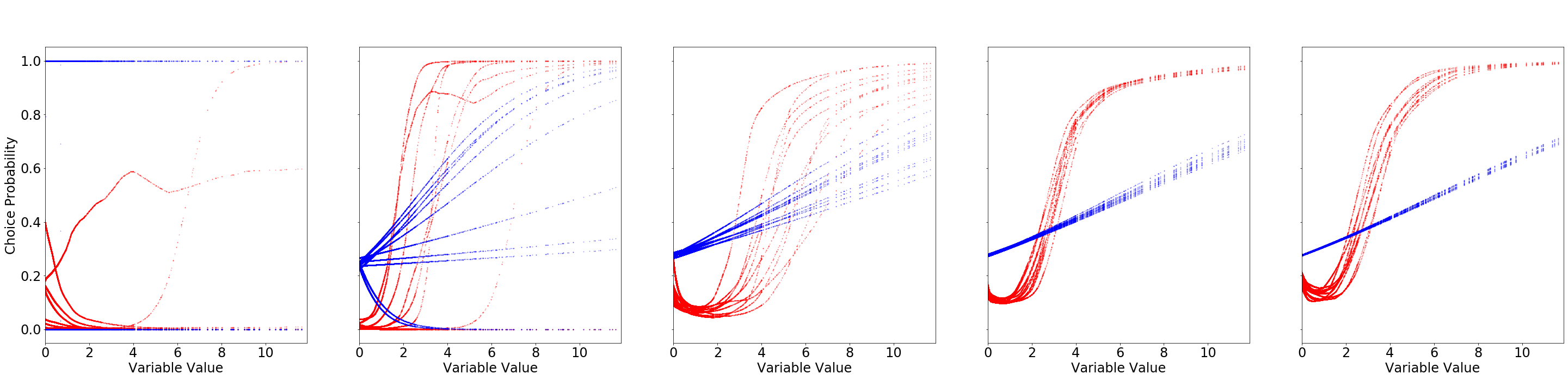}\label{sfig:nhts_mode_dist}}
\caption{NHTS Dataset. Upper row: comparison of DNN and BNL for prediction losses in predicting travel mode choice and trip purposes; lower row: visualizing how choice probabilities change with inputs.}
\label{fig:nhts}
\end{figure}

\noindent
The NHTS dataset is chosen owing to its broad geographical coverage (full U.S.), the large sample size ($780,000$ trips), and the large number of input variables, enabling us to observe the variation of prediction accuracy with varying sample size and input variables. $10\%$ of the NHTS dataset is saved for testing model performance. To form a parallel discussion with our simulations, the NHTS experiment varies sample size (from $100$ to $500,000$) and the number of input variables ($20$ and $50$). The input variables are selected to be the most important ones that determine mode choice and trip purposes. The results are visualized in Figure \ref{fig:nhts}, with the format similar to previous ones but two differences: interpretation losses can no longer be examined since $s^*(x)$ is no longer known and approximation error is no longer available because the theoretical minimum value is unknown either.

Interestingly, Figures from \ref{sfig:nhts_purp_20_var} to \ref{sfig:nhts_purp_50_var} show a pattern that mixes scenarios 1 and 2: BNL outperforms DNN when sample size is around $10^3$, while DNN starts to outperform BNL when sample size is larger than $10^4$. The convergence of BNL is very quick, and it stops at around $10^3$ sample size, while the convergence of DNN is still unclear given that the red curves still have decreasing trend when sample size reaches $500,000$. It again demonstrates that only very large sample size can unleash the full predictive power of DNN, although the $10^4$ sample size is not unattainable in even questionnaire surveys. These results also suggest that handcrafted utility specification has captured certain information, given its better performance when sample size is relatively small, although it does not capture all possible information in the dataset, given its worse performance when sample size is large. Obviously, the approximation error of DNN is smaller than BNL, but the estimation error of DNN does not appear large either.

Figure \ref{sfig:nhts_mode_dist} visualizes how probability of driving changes with trip distance. The results are quite similar to our findings in scenario 2 and 3 in that DNN starts to converge when sample size reaches $10^4$ and its pattern becomes quite stable when sample size equals to $10^5$ or $10^6$. The difference between DNN and BNL implies again that the approximation error exists in BNL. The driving probability functions of DNN and BNL are similar and intuitive in that both are monotonically increasing, while DNN seems to capture more subtlety than BNL: BNL suggests a nearly linear relationship, while DNN describes a relationship with roughly decreasing sensitivity to trip distance changes. This decreasing sensitivity is quite intuitive since people are less likely to change their driving behavior as trip distance is already large enough.

\section{Conclusion and Discussions}
\noindent
This study discusses when and why DNN can be applied to choice analysis, with focuses on answering the non-overfitting and interpretability challenges faced by DNN. A theoretical framework is presented to describe the tradeoff between estimation and approximation errors, and the balance between prediction and interpretation losses. The theory is further demonstrated by using three simulated scenarios and the NHTS dataset, yielding these major findings.

First of all, interpretability can be operationalized by using choice probability functions, owing to the fact that utility comparison and specification naturally exist in DNN and that an accurate estimator $\hat{s}(x)$ of choice probability function enables researchers to extract all the necessary economic information commonly obtained from traditional choice modeling. Our model interpretation is discussed in a way quite different from traditional methods for at least three reasons. (1) The process can be named as prediction-driven interpretation \footnote{It can also be named as post-hoc interpretation, implying that researchers extract information from prediction-driven models after model training. It is debatable whether this approach is the best, since many other alternative approaches exist \cite{Boshi_Velez2017,Ribeiro2016,Lipton2016}}, implying that researchers extract information from DNN after model training even though DNN is designed to maximize prediction accuracy in the first place. This prediction-driven interpretation is intuitive since ``some structure must have been found in DNN, when predictive quality is consistently high'' \cite{Mullainathan2017}. (2) Our interpretation is based on function estimation rather than parameter estimation. It is nearly impossible to evaluate each individual parameter in DNN, so function estimation that focuses on the whole space of the transformed feature in DNN is a more viable way for interpretation. (3) This prediction-driven interpretation approach could automatically learn the underlying utility specification, as shown in our Monte Carlo simulations and the NHTS application. This approach is effective since handcrafted utility specification can rarely capture the full information, and correspondingly, certain power of automatic learning utility specification should always be involved in choice analysis.

Second, the non-overfitting issue can be at least partially addressed by recent progresses in statistical learning theory and demonstrated in our experiments. The estimation error of both prediction and interpretation losses can be bounded by Rademacher complexity of DNN. It is still challenging to provide a clear-cut rule about the correct sample size, since the theoretical development is still on-going and the theory suggests a subtle dynamics between sample size, input dimensions and scale, DNN depth, and norms of each layer in DNN. However, the bottom-line is that researchers do not need to count the number of parameters to bound estimation error of DNN, since the VC dimension based upper bound is too loose for DNN applications. Although sample size requirement is not as large as expected from classical statistical theory, a relatively large sample is still critical for generalizable results from DNN. Our experiments suggests that sample size needs to reach at least $10^4$ for DNN to outperform BNL for typical travel behavior analysis. This requirement of sample size is slightly larger than the common size of questionnaire surveys, but still attainable in practice. In fact, several studies that found DNNs outperforming MNL have sample sizes with a similar magnitude to $10^4$ \cite{XieChi2003,Omrani2015}. However, even when sample size is less than $10^4$, it does not imply that DNN cannot work. In this case, careful regularization methods can and should be used to improve model performance, although we do not focus on regularization much in this study. 

%For choice modeling practice, we have the following suggestions. DNN can be seen as a natural extension of DCMs because of their similarity in terms of utility comparison and specification. DNN is more powerful owing to its function approximation power and automatic feature learning capacity. Practitioners can use DNNs to achieve higher performance than DCMs in terms of both prediction and interpretation, unless the sample size is small ($< 10^4$) or the modelers have a strong belief in the completeness of their prior knowledge.

We believe these insights contribute to the understanding when and why DNN can be used for choice analysis, and they are of both theoretical and practical importance. The theoretical framework can serve as a new foundation for future investigation in choice analysis, since it extends the classical asymptotic data regime ($v/N \rightarrow 0$) to the non-asymptotic data regime, or equivalently, from low-dimensional statistical to high-dimensional statistical tools by using the most recent progresses in statistical learning theory. This extension is important since the non-asymptotic data regime and the high-dimensional statistical tools are becoming increasingly common in practice. Meanwhile, researchers can use the interpretation ideas to generate economic information from DNN-based choice models to achieve the level of interpretability at least the same as traditional choice models, serving for behavioral and policy analysis purposes. However, many important tasks still remain for future studies. Each one of the four quadrants is much deeper and complicated than our discussion. Future studies will need to look into effective regularization methods for small samples, investigate how DNN relates to mixed logit models that have random coefficients, broaden the interpretability concept in a way beyond that framed in traditional choice models, and shed lights on the approximation error part of DNN. Given the richness of machine learning models and the importance of individual decision-making, their intersection will undoubtedly be a fertile research field in the future. 

\section*{Contributions of the Authors}
\noindent
S.W. and J.Z. conceived of the presented idea; S.W. developed the theory and reviewed previous studies; S.W. and Q.W. designed and conducted the experiments; S.W. and N.B. drafted the manuscripts; S.W. derived the analytical proofs. J.Z. supervised this work. All authors discussed the results and contributed to the final manuscript.

\section*{Acknowledgement}
\noindent
We thank Singapore-MIT Alliance for Research and Technology (SMART) for partially funding this research. We thank Sun Rui for his help in verifying the math proofs.

\newpage
\includepdf[pages={1,2,3,4,5}]{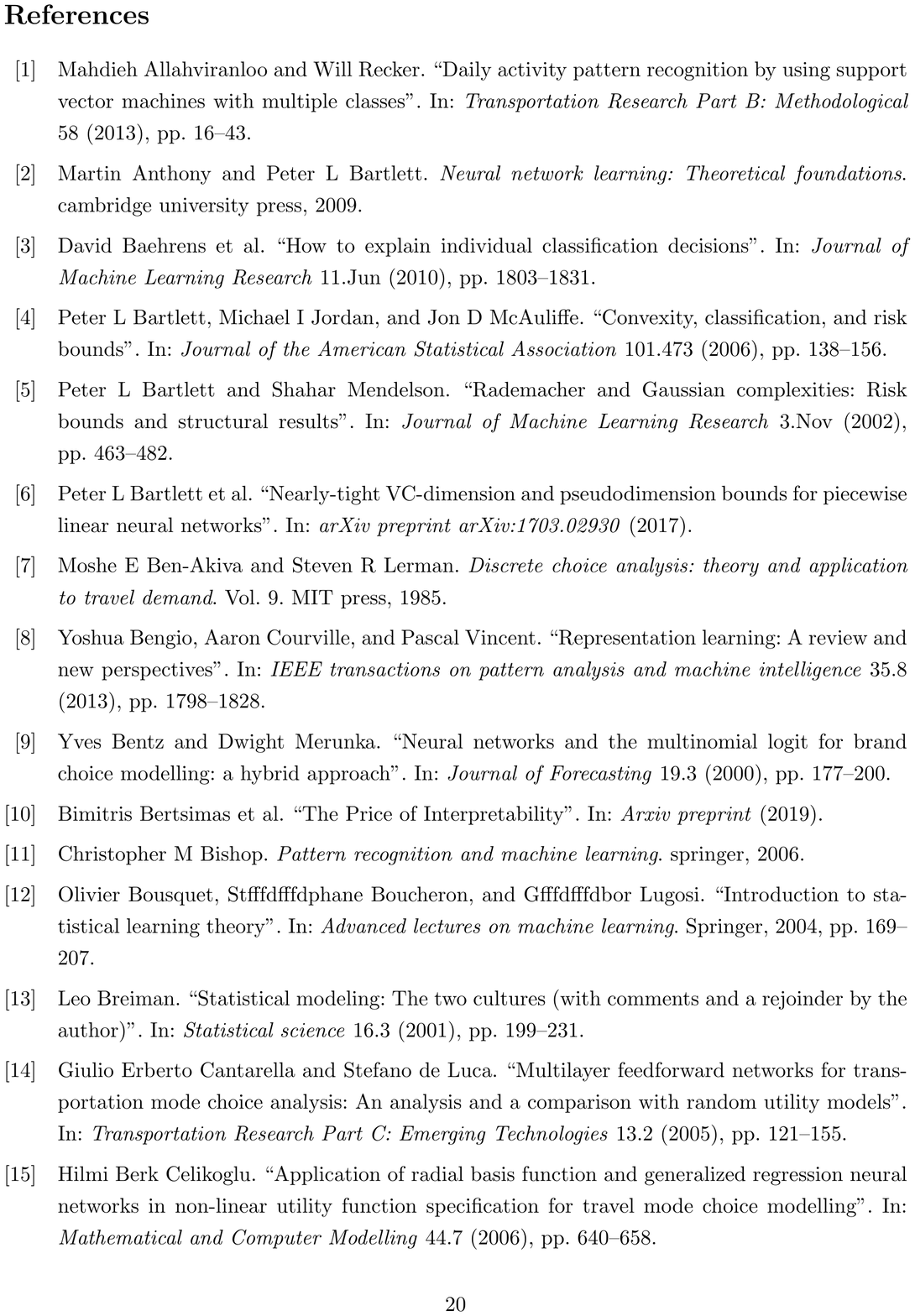}
%\printbibliography

\newpage
\section*{Appendix I: BNL as One Special Case of DNN}
\noindent
Suppose individuals choose between two alternatives $0$ and $1$, which have the utility specifications: $ U_{i0} = V_{i0} + \epsilon_{i0}; \ U_{i1} = V_{i1} + \epsilon_{i1} $, in which $V$ is the deterministic utility and $\epsilon$ is the random utility term. Choice modeling assumes that individuals seek to maximize utility, so the choice probability functions are given by Equation \ref{eq:choice_prob} when $\epsilon$ follows extreme value distribution $EV(0, 1)$.

\begin{equation}
\setlength{\jot}{2pt} \label{eq:choice_prob}
  \begin{aligned}
  P(y_i = 1 | x_i, w) & = \frac{1}{1 + e^{-(V_{i1} - V_{i0})}} \\
  P(y_i = 0 | x_i, w) & = \frac{1}{1 + e^{+(V_{i1} - V_{i0})}} \\
  \end{aligned}
\end{equation}

\noindent
Assuming that attributes relevant to alternatives $0$ and $1$ are $x_{i0}$ and $x_{i1}$, the deterministic utility function with linear specification is

\begin{equation}
\setlength{\jot}{2pt} \label{eq:util_specification_1}
  \begin{aligned}
  V_{i0}(x_{i0}) &= \langle w_0, x_{i0} \rangle \\
  V_{i1}(x_{i1}) &= \langle w_1, x_{i1} \rangle \\
  \end{aligned}
\end{equation}

\noindent
This specification could be more involved by using some transformation $\phi(x)$ (quadratic or log) based on prior knowledge. Hence often the real utility specification could be denoted as 

\begin{equation}
\setlength{\jot}{2pt} \label{eq:util_specification_2}
  \begin{aligned}
  V_{i0}(x_{i0}) &= \langle w_0, \phi(x_{i0}) \rangle \\
  V_{i1}(x_{i1}) &= \langle w_1, \phi(x_{i1}) \rangle \\
  \end{aligned}
\end{equation}

\noindent
This specification is quite close to that in DNN, which is:

\begin{equation}
\setlength{\jot}{2pt} \label{eq:util_specification_3}
  \begin{aligned}
  V_{i0}(x_{i0}) &= \langle w_0, (g_{m-1} ... \circ g_2 \circ g_1)(x_{i0}) \rangle \\
  V_{i1}(x_{i1}) &= \langle w_1, (g_{m-1} ... \circ g_2 \circ g_1)(x_{i1}) \rangle \\
  \end{aligned}
\end{equation}

\noindent
in which $g_j(x) = ReLU(\langle W_j, x \rangle)$. Comparing Equations \ref{eq:util_specification_1}, \ref{eq:util_specification_2}, and \ref{eq:util_specification_3}, it is not hard to see that DNN specification is more general than previous two. A more formal way to demonstrate this point is to use the results from McFadden (1974) \cite{McFadden1974}, which proved that Softmax activation function implicitly implies a random utility maximization with a random utility term that follows Gumbel distribution. By changing the notation of Equation \ref{eq:util_specification_3}, 

\begin{equation}
\setlength{\jot}{2pt} \label{eq:util_specification_4}
  \begin{aligned}
  {\Phi_1(x_i, w)} = V_{i1}(x_{i1})-V_{i0}(x_{i0}) = (g_m \circ ... \circ g_2 \circ g_1)(x_i)
  \end{aligned}
\end{equation}

\noindent
in which $g_m$ is $w_1 - w_0$ and $x_i$ includes all input information. Then Equation \ref{eq:util_specification_4} implies the choice probability of in DNN is:

\begin{equation}
\setlength{\jot}{2pt}
  \begin{aligned}
  \sigma(\Phi_1(x_i, w)) = \frac{1}{1 + e^{-\Phi_1(x_i, w)}}
  \end{aligned}
\end{equation}

\noindent
which is the same as Equation \ref{eq:choice_probability_dnn}.

\section*{Appendix II.A: Proof of Proposition \ref{prop:excess_error_rad_complexity}}
\noindent 
Estimation error can be decomposed:
\begin{flalign}
{\E}_S [L(\hat f) - L(f^*_F)] &= {\E}_S[L(\hat f) - \hat L(\hat f) + \hat L(\hat f) - \hat L(f^*_F) + \hat L(f^*_F) - L(f^*_F)] \\
  &\leq {\E}_S[L(\hat f) - \hat L(\hat f)] \\
  &\leq {\E}_S[ \underset{f \in F}{\sup} \ |L(f) - \hat L(f)|] \label{eq:uniform_deviation} 
%\label{eq:inequality}
\end{flalign}

\noindent
The first inequality holds since (1) $\hat L(\hat f) - \hat L(f^*_F) \leq 0$ due to the definition of $\hat{f}$ and (2) ${\E}_S[\hat L(f^*_F) - L(f^*_F)] = 0$ due to law of large numbers; the second inequality holds since $\hat{f}$ is only one function in $F$ (${\F}_0$ or ${\F}_1$ in this study). The right hand side of Equation \ref{eq:uniform_deviation} above can be further upper bounded by using a technique called symmetrization. Formally, suppose another set of $\{x_i'\}_1^N$ is also generated, following the same distribution as $\{x_i\}_1^N$. Then
\begin{flalign}
{\E}_S \Big[ \underset{f \in F}{\sup} \ \Big| L(f) - \hat L(f) \Big| \Big] &= {\E}_S \Big[ \underset{f \in F}{\sup} \ \Big| {\E}_{x,y}[l(y,f(x))] - \frac{1}{N} \sum_{i=1}^N l(y_i, f(x_i))\Big| \Big] \\ 
	&= {\E}_S \Big[ \underset{f \in F}{\sup} \ \Big| \frac{1}{N} \sum_{i=1}^N {\E}_{x'} l(y,f(x_i')) - \frac{1}{N} \sum_{i=1}^N l(y_i, f(x_i))\Big| \Big] \\ 
	&\leq {\E}_{S,S'} \Big[ \underset{f \in F}{\sup} \ \Big| \frac{1}{N} \sum_{i=1}^N l(y,f(x_i')) - \frac{1}{N} \sum_{i=1}^N l(y_i, f(x_i))\Big| \Big] \\ 
	&= {\E}_{S,S'} \Big[ \underset{f \in F}{\sup} \ \Big| \frac{1}{N} \sum_{i=1}^N \epsilon_i (l(y,f(x_i')) - l(y_i, f(x_i))\Big| \Big] \\ 
	&\leq {\E}_{S,S'} \Big[ \underset{f \in F}{\sup} \ \Big| \frac{1}{N} \sum_{i=1}^N \epsilon_i l(y,f(x_i') \Big| + \Big| \frac{1}{N} \sum_{i=1}^N \epsilon_i l(y_i, f(x_i) \Big| \Big] \\
	&\leq 2 {\E}_S \hat{\R}_n(l \circ {\F} |_S)
\end{flalign}

\noindent
The first line uses the definition of $L$ and $\hat{L}$; the second line uses the symmetrization technique by which ${\E}_{x,y}$ is replaced by an average of another sample $\frac{1}{N} \sum_{i=1}^N {\E}_{x'} l(y,f(x_i'))$; the third line uses $\E \ \sup \geq \sup \ \E$ and uses $S'$ to denote the new sample $\{x' \}_1^N$; the fourth line adds the Rademacher random variable ${\epsilon}_i$ due to the symmetry of $S$ and $S'$; the fifth line uses the fact $\sup |A + B| \leq \sup |A| + \sup |B|$; and the last line is the definition of Rademacher complexity. 

\section*{Appendix II.B: Proof of Proposition \ref{prop:excess_error_rad_complexity_prediction}}
\begin{definition} \label{def:ramp_loss}
Ramp loss function is defined as 
\begin{flalign}
\phi(s) = 
\Bigg\{
  \begin{tabular}{ccc}
  $1$ & $s \leq 0$ \\
  $1 - s/\gamma$ & $0 < s < \gamma$ \\
  $0$ & $s \geq \gamma$
  \end{tabular}
%\right \}
%\label{eq:present_discounting}
\end{flalign}
Associated error function is
\begin{flalign}
L_{\phi} = \E[\phi(s)]
\end{flalign}
\end{definition}

\begin{definition} \label{def:gamma_margin}
$\gamma$-margin loss function is defined as
\begin{flalign}
\I\{y \Phi(x) \leq \gamma \}
\end{flalign}
Associated error function is
\begin{flalign}
L_{\gamma} = \E[\I\{y \Phi(x) \leq \gamma \}]
\end{flalign}
\end{definition}

\noindent
$L_{\phi}$ is an example of \textit{surrogate loss} functions for $L_{0/1}$. It is a surrogate loss since $L_{\phi}$ is designed to (1) upper bound $L_{0/1}$ and (2) be L-Lipschitz so that the contraction inequality can be applied. The Lipschitz constant of $L_{\phi}$ is $1/\gamma$. By design, three error functions are related:
\begin{equation}
L_{0/1} \leq L_{\phi} \leq L_{\gamma}
\end{equation}

\noindent
Therefore, the estimation error measured by prediction error $L_{0/1}$ can be upper bounded 
\begin{flalign}
{\E}_S[L_{0/1}(\hat f) - \hat L_{\gamma}(\hat f)] \leq {\E}_S[L_{\phi}(\hat f) - \hat L_{\phi}(\hat f)]
\label{eq:surrogate_bound}
\end{flalign}

The right hand side of Equation \ref{eq:surrogate_bound} can be upper bounded by using Proposition \ref{prop:excess_error_rad_complexity} and contraction inequality.
\begin{flalign}
{\E}_S[L_{\phi}(\hat f) - \hat L_{\phi}(\hat f)] &\leq {\E}_S \ \underset{f \in {\F}_1}{\sup} \ |L_{\phi}(f) - \hat L_{\phi}(f)| \\
  &= {\E}_S \ \underset{f \in {\F}_1}{\sup} \ |\E[\phi(f)] - \frac{1}{N} \sum_{i=1}^N \phi(f(x_i))| \\
  &= {\E}_{S,\epsilon} \ \underset{f \in {\F}_1}{\sup} \ \frac{2}{N} \sum_{i=1}^N |\epsilon_i \phi(f(x_i))| \\
  &\leq \frac{2}{\gamma} \times {\E}_{S,\epsilon} \ \underset{f \in {\F}_1}{\sup} \  \frac{1}{N} \sum_{i=1}^N |\epsilon_i f(x_i)| \\
  &= \frac{2}{\gamma} {\E}_{S,\epsilon} \hat{\R}_n({\F}_1|_S)
\end{flalign}

\noindent
The first inequality holds due to the $\sup$ operator; the second line uses the definitions of ramp cost functions; the third line used Proposition \ref{prop:excess_error_rad_complexity}; the fourth line used contraction inequality \cite{Ledoux2013}; and the last line used the definition of empirical Rademacher complexity. Using Equation \ref{eq:surrogate_bound}, it implies

\begin{flalign}
{\E}_S[L_{0/1}(\hat f) - \hat L_{\gamma}(\hat f)] \leq \frac{2}{\gamma} {\E}_{S,\epsilon} \hat{\R}_n({\F}_1|_S)
\end{flalign}

\noindent Therefore, the $L_{0/1}(\hat{f})$ can be upper bounded by empirical $\gamma$-margin loss plus Rademacher complexity. $\hat L_{\gamma}(\hat f)$ can be empirically computed, so a valid upper bound exists for $L_{0/1}(\hat{f})$. However, the unresolved question is whether DNN automatically finds a maximum margin similar to SVM, so that the $L_{0/1}(\hat{f})$ is bounded well. It is still an on-going research field \cite{Poggio2018_2,Poggio2018_1,Soudry2018}. $\square$ \\

\section*{Appendix II.C: Proof of Proposition \ref{prop:excess_error_rad_complexity_interpretation}}
\begin{definition} \label{def:mse}
Mean squared error (MSE) is defined as
\begin{flalign}
L_{mse}(s) = {\E}_{x,y} [(y - s(x))^2]
\end{flalign}
The corresponding empirical mean squared error is defined as 
\begin{flalign}
\hat L_{mse}(s) = \frac{1}{N} \sum_{i=1}^N (y_i - s(x_i))^2
\end{flalign}
\end{definition}

\begin{lemma} \label{lemma:mse_estimation_error_int}
Estimation error for interpretation equals to that of MSE.
\begin{flalign}
{\E}_S [L_{mse}(\hat{s}) - L_{mse}(s^*_F))] = {\E}_S [L_{e}(\hat{s}) - L_{e}(s^*_F))] 
\end{flalign}
\end{lemma}

\noindent
\textbf{Proof of Lemma \ref{lemma:mse_estimation_error_int}.} Since $y$ is sampled as a Bernoulli random variable with probability $s^*(x)$, $E[y|x] = s^*(x)$. 

\begin{flalign}
{\E}_{S,x,y}[(\hat{s}(x) - y)^2] &= {\E}_{S,x,y}((\hat{s}(x) - s^*(x) + s^*(x) - y)^2) \\
	&= {\E}_{S,x,y}[((\hat{s}(x) - s^*(x))^2 + 2(\hat{s}(x) - s^*(x))(s^*(x) - y) + (s^*(x) - y)^2)] \\
	&= {\E}_{S,x,y}[(\hat{s}(x) - s^*(x))^2] + {\E}_{x,y}[(s^*(x) - y)^2)] + 2{\E}_{S,x,y}[(\hat{s}(x) - s^*(x))(s^*(x) - y)] \\
	&= {\E}_{S,x,y}[(\hat{s}(x) - s^*(x))^2] + {\E}_{x,y}[(s^*(x) - y)^2)] + 2{\E}_{x} \big[ {\E}_{S,y}[(\hat{s}(x) - s^*(x))(s^*(x) - y)|x] \big] \\
	&= {\E}_{S,x,y}[(\hat{s}(x) - s^*(x))^2] + {\E}_{x,y}[(s^*(x) - y)^2)] + 2{\E}_{x} \big[ {\E}_{S}[(\hat{s}(x) - s^*(x))|x] {\E}_{y}[(s^*(x) - y)|x] \big] \\
	&= {\E}_{S,x,y}[(\hat{s}(x) - s^*(x))^2] + {\E}_{x,y}[(s^*(x) - y)^2)] 
\end{flalign}

\noindent
The fourth equality uses Law of Iterated Expectation; the fifth uses the conditional independence $S \perp y | x$; the lase one uses $E[y|x] = s^*(x)$. With very similar process, we could show

\begin{flalign}
{\E}_{x,y}[(y - s^*_F(x))^2] &= {\E}_{x,y}[(y - s^*(x) + s^*(x) - s^*_F(x))^2] \\
	&= {\E}_{x,y}[(y - s^*(x))^2] + {\E}_{x,y}[(s^*(x) - s^*_F(x))^2] + 2{\E}_{x,y}[(y - s^*(x))(s^*(x) - s^*_F(x))] \\
	&= {\E}_{x,y}[(y - s^*(x))^2] + {\E}_{x,y}[(s^*(x) - s^*_F(x))^2] + 2{\E}_{x} \big[ (s^*(x) - s^*_F(x)){\E}_{y}[y - s^*(x)|x]\big] \\
	&= {\E}_{x,y}[(y - s^*(x))^2] + {\E}_{x,y}[(s^*(x) - s^*_F(x))^2]
\end{flalign}

\noindent Combining the two equations above implies
\begin{flalign}
{\E}_{x,y}[(s^*(x) - y)^2)] &= {\E}_{S,x,y}[(\hat{s}(x) - y)^2] - {\E}_{S,x,y}[(\hat{s}(x) - s^*(x))^2] \\
	&= {\E}_{x,y}[(y - s^*_F(x))^2] - {\E}_{x,y}[(s^*(x) - s^*_F(x))^2]
\end{flalign}

\noindent By changing the notation, it implies
\begin{flalign}
	{\E}_S [L_{mse}(\hat{s}) - L_{mse}(s^*_F))] = {\E}_S [L_{e}(\hat{s}) - L_{e}(s^*_F))] 
\end{flalign}

\noindent \textbf{Proof of Proposition \ref{prop:excess_error_rad_complexity_interpretation}.} Lemma \ref{lemma:mse_estimation_error_int} shows that the estimation error on function estimation is the same as the one on MSE. Hence we will provide an upper bound on the MSE by using Proposition \ref{prop:excess_error_rad_complexity}. Formally,

\begin{flalign}
{\E}_S [L_{mse}(\hat{s}) - L_{mse}(s^*_F))] &\leq 2 {\E}_S [\hat{R}_n(l \circ \F |_S)] \\
	&\leq 4 {\E}_S [\hat{R}_n(\F |_S)] 
\end{flalign}

\noindent
The first inequality uses Proposition \ref{prop:excess_error_rad_complexity}; the second uses contraction inequality and the fact that squared loss here is bounded between $[0,1]$ and that its Lipschitz constant is at most two. $\square$

\section*{Appendix II.D: Proof of Proposition \ref{prop:rad_complexity_dnn_contraction}}
\noindent
The proof is an iterative process going through layer by layer. Suppose for layer $j$ of DNN, the mapping is

$$ F_j = \{f: x \rightarrow \sum_{t=1}^{d_{j-1}} w_t \sigma(f_{t}(x)); f_{t} \in F_{j-1}, ||w||_1 \leq M(j) \} $$

\noindent
Then the Rademacher complexity of $F_j$ can be represented by that of $F_{j-1}$.

\begin{flalign}
N \hat{\R}_n(F_j|_S) &= \E_{\epsilon} \underset{f_j \in F_j}{\sup} \ \Big| \sum_{i=1}^N \epsilon_i f(x_i) \Big| \\
	&= \E_{\epsilon} \underset{f_j \in F_j}{\sup} \ \Big| \sum_{i=1}^N \epsilon_i \sum_{t = 1}^{d_{j-1}} w_t \sigma(f_{t}(x_i)) \Big| \\
  &= {\E}_{\epsilon} \underset{\substack{||w||_1 \leq M(j) \\ f_{t} \in F_{j-1}}}{\sup} \ \Big| \sum_{t = 1}^{d_{j-1}} w_t \sum_{i=1}^N \epsilon_i \sigma(f_{t}(x_i)) \Big| \\
  &= 2{\E}_{\epsilon} \underset{\substack{||w||_1 \leq M(j) \\ f_{t} \in F_{j-1}}}{\sup} \ \sum_{t = 1}^{d_{j-1}} w_t \sum_{i=1}^N \epsilon_i \sigma(f_{t}(x_i)) \\
  &= 2 M(j) {\E}_{\epsilon} \underset{f_{t} \in F_{j-1}}{\sup} \ \underset{t}{\max} \Big| \sum_{i=1}^N \epsilon_i \sigma(f_{t}(x_i)) \Big| \\
  &= 2 M(j) {\E}_{\epsilon} \underset{f_{t} \in F_{j-1}}{\sup} \ \Big| \sum_{i=1}^N \epsilon_i \sigma(f_{t}(x_i)) \Big| \\
  &\leq 2 M(j) {\E}_{\epsilon} \underset{f_{t} \in F_{j-1}}{\sup} \ \Big| \sum_{i=1}^N \epsilon_i f_{t}(x_i) \Big| \\
  &\leq 2 M(j) N \hat{\R}_n(F_{j-1}|_S)
\label{eq:dnn_rad}
\end{flalign}
\noindent
which implies this iterative formula for DNN:
\begin{flalign}
\hat{\R}_n(F_j|_S) = 2 M(j) \hat{\R}_n(F_{j-1}|_S)
\end{flalign}

The remaining question is about the Rademacher complexity of layer $0$, which is a linear transformation $F_0 = \{x \rightarrow \langle w, x \rangle: w \in B^d_1 \}$ with normalized input $X$.
\begin{flalign}
\hat{\R}_n(F_{0}|_S) \leq \sqrt{\frac{\log d_0}{N}}
\end{flalign}

\noindent
Combining the equations above, Rademacher complexity of DNN can be proved as:
\begin{flalign}
\hat{\R}_n({\F}_{1}|_S) \lesssim \frac{\sqrt{\log d_0} \times \prod_{j=1}^D 2M(j)}{\sqrt{N}} 
\end{flalign}

\noindent
Note that here the Rademacher complexity has the $2^D$ factor. With more involved technique, a tighter upper bound could be proved as

\begin{flalign}
\hat{\R}_n({\F}_{1}|_S) \lesssim \frac{\sqrt{\log d_0} \times (\sqrt{2 \log(D)} + 1) \prod_{j=1}^D M_F(j)}{\sqrt{N}}
\end{flalign}

This result can be found in in Golowich et al. (2017) \cite{Golowich2017}, with slight differences. The key steps of the proof we presented here can be found in Bartlett and Mendelson (2002) \cite{Bartlett2002}. Other relevant work can be found in \cite{Anthony2009} and \cite{Neyshabur2015}. 

\section*{Appendix II.E. Proof of Proposition \ref{prop:rad_complexity_dnn_vc}}
\noindent
Since VC dimension is only used as a benchmark, we will demonstrate a simple proof that upper bound the estimation error by $O(\sqrt{\frac{v \log(N+1)}{N}})$ for binary output. Using Lemma 4.14 from Wainwright (2019) \cite{Wainwright2019}

\begin{flalign}
\hat{\R}_n(l \circ {\F}_1 |_S) \leq 4 \sqrt{\frac{v \log(N + 1)}{N}}
\end{flalign}

\noindent
Note that $\log(N+1)$ is much smaller than $v$ and $N$. This upper bound can be simplified to 
\begin{flalign}
O(\sqrt{\frac{v}{N}})
\end{flalign}

\noindent
which is similar to the traditional wisdom of examining the ratio between number of parameters and number of observations, since $v$ is the same as parameter numbers in generalized linear models. For DNN, the tightest possible VC dimension can be found in \cite{Bartlett2017}, which is $v = O(TD \log(T))$ with $T$ denoting the total number of coefficients and $D$ the depth of DNN. This $O(\sqrt{\frac{v}{N}})$ can also be used for the $\hat{s}(x)$ case. But we won't discuss details here. Readers could refer to \cite{Vapnik1999,Vapnik2013,Von_Luxburg2011,Wainwright2019} for details. 

\section*{Appendix III: Further Results in Experiments}
\begin{figure}[t!]
\centering
\subfloat[Choice Probability Curves (50 Var)]{\includegraphics[width=\linewidth]{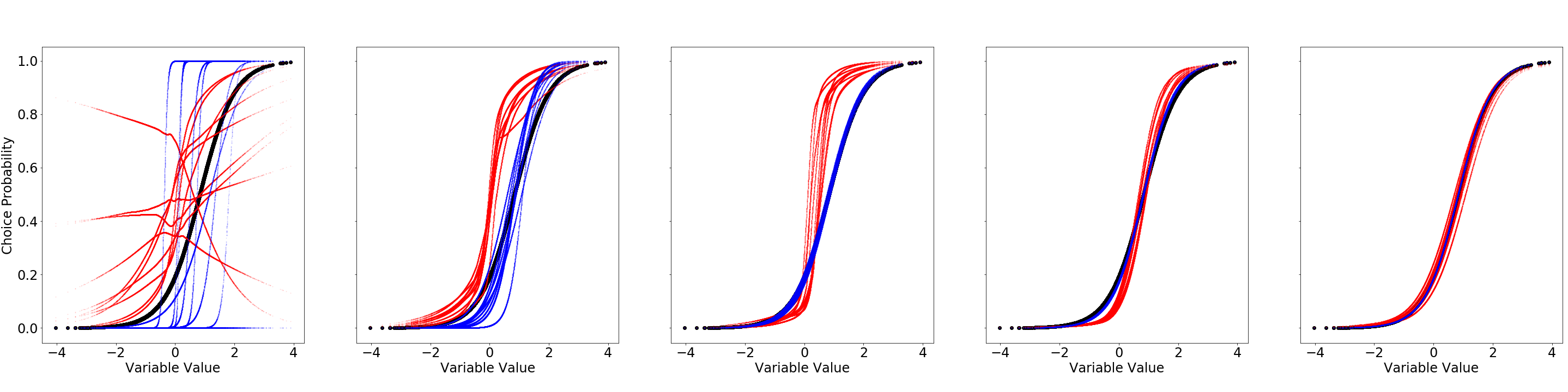}\label{sfig:s1_chprob_50_var}} \\
\subfloat[Choice Probability Curves (50 Var)]{\includegraphics[width=\linewidth]{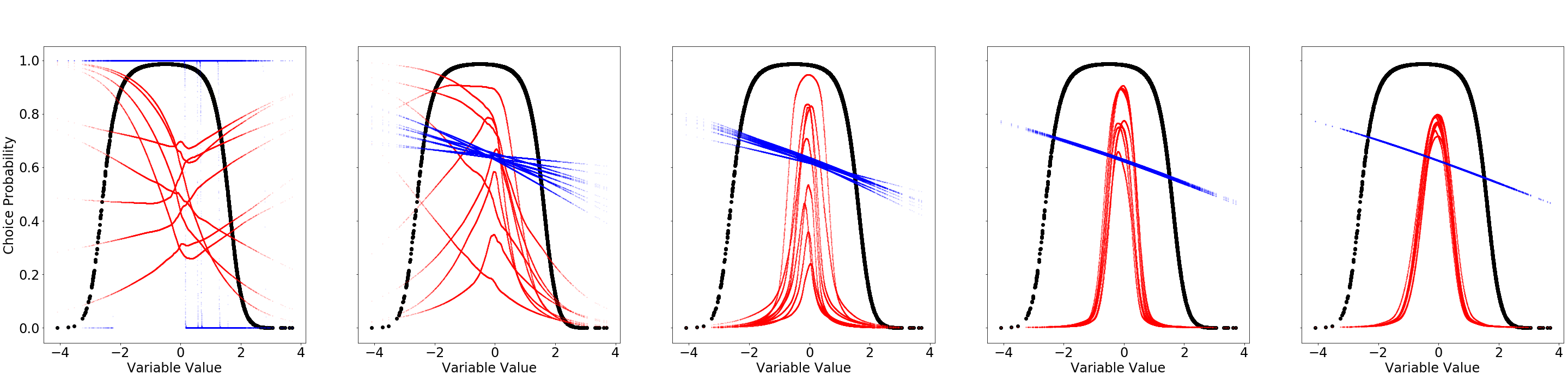}\label{sfig:s2_chprob_50_var}} \\
\subfloat[Choice Probability Curves (50 Var)]{\includegraphics[width=\linewidth]{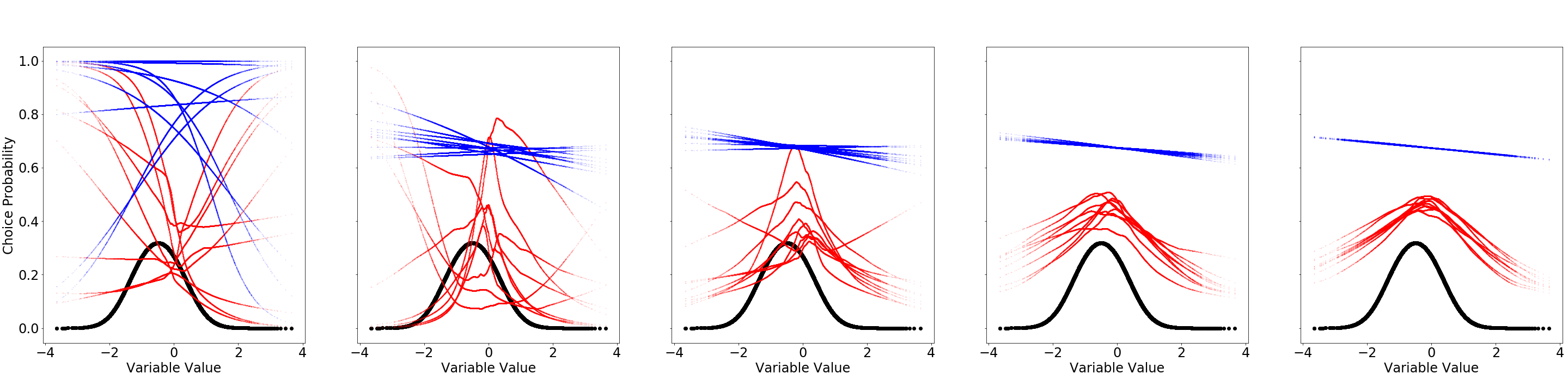}\label{sfig:s3_chprob_50_var}} \\
\caption{Scenario 1-3. From Left to Right: Sample Size 100, 1000, 10000, 100000, 1000000}
\label{fig:sce_chprob}
\end{figure}

\noindent
Results about $50$ variables are included in Figure \ref{fig:sce_chprob}

@article{ZhouBolei2014,
   Author = {Zhou, Bolei and Khosla, Aditya and Lapedriza, Agata and Oliva, Aude and Torralba, Antonio},
   Title = {Object detectors emerge in deep scene cnns},
   Journal = {arXiv preprint arXiv:1412.6856},
   Year = {2014} }

@article{Zegras2010,
   Author = {Zegras, Christopher},
   Title = {The built environment and motor vehicle ownership and use: Evidence from Santiago de Chile},
   Journal = {Urban Studies},
   Volume = {47},
   Number = {8},
   Pages = {1793-1817},
   Year = {2010} }

@article{XieChi2003,
   Author = {Xie, Chi and Lu, Jinyang and Parkany, Emily},
   Title = {Work travel mode choice modeling with data mining: decision trees and neural networks},
   Journal = {Transportation Research Record: Journal of the Transportation Research Board},
   Number = {1854},
   Pages = {50-61},
   Year = {2003} }

@incollection{Von_Luxburg2011,
   Author = {Von Luxburg, Ulrike and Schölkopf, Bernhard},
   Title = {Statistical learning theory: Models, concepts, and results},
   BookTitle = {Handbook of the History of Logic},
   Publisher = {Elsevier},
   Volume = {10},
   Pages = {651-706},
   Year = {2011} }

@book{Vapnik2013,
   Author = {Vapnik, Vladimir},
   Title = {The nature of statistical learning theory},
   Publisher = {Springer science and business media},
   Year = {2013} }

@article{Vapnik1999,
   Author = {Vapnik, Vladimir Naumovich},
   Title = {An overview of statistical learning theory},
   Journal = {IEEE transactions on neural networks},
   Volume = {10},
   Number = {5},
   Pages = {988-999},
   Year = {1999} }

@book{Train2009,
   Author = {Train, Kenneth E},
   Title = {Discrete choice methods with simulation},
   Publisher = {Cambridge university press},
   Year = {2009} }

@article{Train1980,
   Author = {Train, Kenneth},
   Title = {A structured logit model of auto ownership and mode choice},
   Journal = {The Review of Economic Studies},
   Volume = {47},
   Number = {2},
   Pages = {357-370},
   Year = {1980} }

@article{Sontag1998,
   Author = {Sontag, Eduardo D},
   Title = {VC dimension of neural networks},
   Journal = {NATO ASI Series F Computer and Systems Sciences},
   Volume = {168},
   Pages = {69-96},
   Year = {1998} }

@inproceedings{Ribeiro2016,
   Author = {Ribeiro, Marco Tulio and Singh, Sameer and Guestrin, Carlos},
   Title = {Why should i trust you?: Explaining the predictions of any classifier},
   BookTitle = {Proceedings of the 22nd ACM SIGKDD International Conference on Knowledge Discovery and Data Mining},
   Publisher = {ACM},
   Pages = {1135-1144},
   Year = {2016} }

@inproceedings{Paredes2017,
   Author = {Paredes, Miguel and Hemberg, Erik and O'Reilly, Una-May and Zegras, Chris},
   Title = {Machine learning or discrete choice models for car ownership demand estimation and prediction?},
   BookTitle = {Models and Technologies for Intelligent Transportation Systems (MT-ITS), 2017 5th IEEE International Conference on},
   Publisher = {IEEE},
   Pages = {780-785},
   Year = {2017} }

@book{Ortuzar2011,
   Author = {De Dios Ortuzar, Juan and Willumsen, Luis G},
   Title = {Modelling transport},
   Publisher = {John Wiley and Sons},
   Year = {2011} }

@article{Omrani2015,
   Author = {Omrani, Hichem},
   Title = {Predicting travel mode of individuals by machine learning},
   Journal = {Transportation Research Procedia},
   Volume = {10},
   Pages = {840-849},
   Year = {2015} }

@article{Mullainathan2017,
   Author = {Mullainathan, Sendhil and Spiess, Jann},
   Title = {Machine learning: an applied econometric approach},
   Journal = {Journal of Economic Perspectives},
   Volume = {31},
   Number = {2},
   Pages = {87-106},
   Year = {2017} }

@article{McFadden1974,
   Author = {McFadden, Daniel},
   Title = {Conditional logit analysis of qualitative choice behavior},
   Year = {1974} }

@article{Lipton2016,
   Author = {Lipton, Zachary C},
   Title = {The mythos of model interpretability},
   Journal = {arXiv preprint arXiv:1606.03490},
   Year = {2016} }

@article{LeCun2015,
   Author = {LeCun, Yann and Bengio, Yoshua and Hinton, Geoffrey},
   Title = {Deep learning},
   Journal = {Nature},
   Volume = {521},
   Number = {7553},
   Pages = {436-444},
   Year = {2015} }

@inproceedings{Krizhevsky2012,
   Author = {Krizhevsky, Alex and Sutskever, Ilya and Hinton, Geoffrey E},
   Title = {Imagenet classification with deep convolutional neural networks},
   BookTitle = {Advances in neural information processing systems},
   Pages = {1097-1105},
   Year = {2012} }

@article{Kotsiantis2007,
   Author = {Kotsiantis, Sotiris B and Zaharakis, I and Pintelas, P},
   Title = {Supervised machine learning: A review of classification techniques},
   Journal = {Emerging artificial intelligence applications in computer engineering},
   Volume = {160},
   Pages = {3-24},
   Year = {2007} }

@article{Kingma2014,
   Author = {Kingma, Diederik P and Ba, Jimmy},
   Title = {Adam: A method for stochastic optimization},
   Journal = {arXiv preprint arXiv:1412.6980},
   Year = {2014} }

@article{Karlaftis2011,
   Author = {Karlaftis, Matthew G and Vlahogianni, Eleni I},
   Title = {Statistical methods versus neural networks in transportation research: Differences, similarities and some insights},
   Journal = {Transportation Research Part C: Emerging Technologies},
   Volume = {19},
   Number = {3},
   Pages = {387-399},
   Year = {2011} }

@article{Hornik1989,
   Author = {Hornik, Kurt and Stinchcombe, Maxwell and White, Halbert},
   Title = {Multilayer feedforward networks are universal approximators},
   Journal = {Neural networks},
   Volume = {2},
   Number = {5},
   Pages = {359-366},
   Year = {1989} }

@article{Hinton2015,
   Author = {Hinton, Geoffrey and Vinyals, Oriol and Dean, Jeff},
   Title = {Distilling the knowledge in a neural network},
   Journal = {arXiv preprint arXiv:1503.02531},
   Year = {2015} }

@inproceedings{He2015,
   Author = {He, Kaiming and Zhang, Xiangyu and Ren, Shaoqing and Sun, Jian},
   Title = {Delving deep into rectifiers: Surpassing human-level performance on imagenet classification},
   BookTitle = {Proceedings of the IEEE international conference on computer vision},
   Pages = {1026-1034},
   Year = {2015} }

@article{Hagenauer2017,
   Author = {Hagenauer, Julian and Helbich, Marco},
   Title = {A comparative study of machine learning classifiers for modeling travel mode choice},
   Journal = {Expert Systems with Applications},
   Volume = {78},
   Pages = {273-282},
   Year = {2017} }

@book{Goodfellow2016,
   Author = {Goodfellow, Ian and Bengio, Yoshua and Courville, Aaron and Bengio, Yoshua},
   Title = {Deep learning},
   Publisher = {MIT press Cambridge},
   Volume = {1},
   Year = {2016} }

@article{Glaeser2018,
   Author = {Glaeser, Edward L and Kominers, Scott Duke and Luca, Michael and Naik, Nikhil},
   Title = {Big data and big cities: The promises and limitations of improved measures of urban life},
   Journal = {Economic Inquiry},
   Volume = {56},
   Number = {1},
   Pages = {114-137},
   Year = {2018} }

@article{Fernandez2014,
   Author = {Fernández-Delgado, Manuel and Cernadas, Eva and Barro, Senén and Amorim, Dinani},
   Title = {Do we need hundreds of classifiers to solve real world classification problems},
   Journal = {Journal of Machine Learning Research},
   Volume = {15},
   Number = {1},
   Pages = {3133-3181},
   Year = {2014} }

@techreport{Cohen2016,
   Author = {Cohen, Jonathan D and Ericson, Keith Marzilli and Laibson, David and White, John Myles},
   Title = {Measuring time preferences},
   Institution = {National Bureau of Economic Research},
   Year = {2016} }

@article{Cervero1997_3d,
   Author = {Cervero, Robert and Kockelman, Kara},
   Title = {Travel demand and the 3Ds: density, diversity, and design},
   Journal = {Transportation Research Part D: Transport and Environment},
   Volume = {2},
   Number = {3},
   Pages = {199-219},
   Year = {1997} }

@article{Celikoglu2006,
   Author = {Celikoglu, Hilmi Berk},
   Title = {Application of radial basis function and generalized regression neural networks in non-linear utility function specification for travel mode choice modelling},
   Journal = {Mathematical and Computer Modelling},
   Volume = {44},
   Number = {7},
   Pages = {640-658},
   Year = {2006} }

@article{Boshi_Velez2017,
   Author = {Doshi-Velez, Finale and Kim, Been},
   Title = {Towards a rigorous science of interpretable machine learning},
   Year = {2017} }

@book{Bishop2006,
   Author = {Bishop, Christopher M},
   Title = {Pattern recognition and machine learning},
   Publisher = {springer},
   Year = {2006} }

@article{Bengio2013,
   Author = {Bengio, Yoshua and Courville, Aaron and Vincent, Pascal},
   Title = {Representation learning: A review and new perspectives},
   Journal = {IEEE transactions on pattern analysis and machine intelligence},
   Volume = {35},
   Number = {8},
   Pages = {1798-1828},
   Year = {2013} }

@book{Ben_Akiva1985,
   Author = {Ben-Akiva, Moshe E and Lerman, Steven R},
   Title = {Discrete choice analysis: theory and application to travel demand},
   Publisher = {MIT press},
   Volume = {9},
   Year = {1985} }

@inproceedings{HeKaiming2016, 
   Author = {He, Kaiming and Zhang, Xiangyu and Ren, Shaoqing and Sun, Jian},
   Title = {Deep residual learning for image recognition},
   BookTitle = {Proceedings of the IEEE conference on computer vision and pattern recognition},
   Pages = {770-778},
      Year = {2016} }

@article{Cantarella2005,
   Author = {Cantarella, Giulio Erberto and de Luca, Stefano},
   Title = {Multilayer feedforward networks for transportation mode choice analysis: An analysis and a comparison with random utility models},
   Journal = {Transportation Research Part C: Emerging Technologies},
   Volume = {13},
   Number = {2},
   Pages = {121-155},
      Year = {2005} }

@article{Polson2017,
   Author = {Polson, Nicholas G and Sokolov, Vadim O},
   Title = {Deep learning for short-term traffic flow prediction},
   Journal = {Transportation Research Part C: Emerging Technologies},
   Volume = {79},
   Pages = {1-17},
      Year = {2017} }

@article{WuYuankai2018,
   Author = {Wu, Yuankai and Tan, Huachun and Qin, Lingqiao and Ran, Bin and Jiang, Zhuxi},
   Title = {A hybrid deep learning based traffic flow prediction method and its understanding},
   Journal = {Transportation Research Part C: Emerging Technologies},
   Volume = {90},
   Pages = {166-180},
      Year = {2018} }

@techreport{Qianli2018,
   Author = {Liao, Qianli and Poggio, Tomaso},
   Title = {When Is Handcrafting Not a Curse?},
      Year = {2018} }

@article{Bentz2000,
   Author = {Bentz, Yves and Merunka, Dwight},
   Title = {Neural networks and the multinomial logit for brand choice modelling: a hybrid approach},
   Journal = {Journal of Forecasting},
   Volume = {19},
   Number = {3},
   Pages = {177-200},
      Year = {2000} }

@article{Montavon2018, 
   Author = {Montavon, Gregoire and Samek, Wojciech and Muller, Klaus-Robert},
   Title = {Methods for interpreting and understanding deep neural networks},
   Journal = {Digital Signal Processing},
   Volume = {73},
   Pages = {1-15},
      Year = {2018}}

@book{Anthony2009,
   Author = {Anthony, Martin and Bartlett, Peter L},
   Title = {Neural network learning: Theoretical foundations},
   Publisher = {cambridge university press},
      Year = {2009}}

@book{Wainwright2019,
   Author = {Wainwright, Martin J},
   Title = {High-dimensional statistics: A non-asymptotic viewpoint},
   Publisher = {Cambridge University Press},
   Volume = {48},
      Year = {2019}}

@article{Bartlett2002,
   Author = {Bartlett, Peter L and Mendelson, Shahar},
   Title = {Rademacher and Gaussian complexities: Risk bounds and structural results},
   Journal = {Journal of Machine Learning Research},
   Volume = {3},
   Number = {Nov},
   Pages = {463-482},
      Year = {2002}}

@article{Bartlett2006,
   Author = {Bartlett, Peter L and Jordan, Michael I and McAuliffe, Jon D},
   Title = {Convexity, classification, and risk bounds},
   Journal = {Journal of the American Statistical Association},
   Volume = {101},
   Number = {473},
   Pages = {138-156},
      Year = {2006} }

@article{Bartlett2017,
   Author = {Bartlett, Peter L and Harvey, Nick and Liaw, Chris and Mehrabian, Abbas},
   Title = {Nearly-tight VC-dimension and pseudodimension bounds for piecewise linear neural networks},
   Journal = {arXiv preprint arXiv:1703.02930},
      Year = {2017} }

@article{Golowich2017,
   Author = {Golowich, Noah and Rakhlin, Alexander and Shamir, Ohad},
   Title = {Size-independent sample complexity of neural networks},
   Journal = {arXiv preprint arXiv:1712.06541},
      Year = {2017} }

@inproceedings{Neyshabur2015,
   Author = {Neyshabur, Behnam and Tomioka, Ryota and Srebro, Nathan},
   Title = {Norm-based capacity control in neural networks},
   BookTitle = {Conference on Learning Theory},
   Pages = {1376-1401},
      Year = {2015} }

@techreport{Poggio2018_1,
   Author = {Poggio, Tomaso and Kawaguchi, Kenji and Liao, Qianli and Miranda, Brando and Rosasco, Lorenzo and Boix, Xavier and Hidary, Jack and Mhaskar, Hrushikesh},
   Title = {Theory of deep learning iii: the non-overfitting puzzle},
   Institution = {Technical report, Technical report, CBMM memo 073},
      Year = {2018}}

@article{Poggio2018_2,
   Author = {Poggio, Tomaso and Liao, Qianli and Miranda, Brando and Banburski, Andrzej and Boix, Xavier and Hidary, Jack},
   Title = {Theory IIIb: Generalization in deep networks},
   Journal = {arXiv preprint arXiv:1806.11379},
      Year = {2018}}

@article{Soudry2018,
   Author = {Soudry, Daniel and Hoffer, Elad and Nacson, Mor Shpigel and Gunasekar, Suriya and Srebro, Nathan},
   Title = {The implicit bias of gradient descent on separable data},
   Journal = {The Journal of Machine Learning Research},
   Volume = {19},
   Number = {1},
   Pages = {2822-2878},
      Year = {2018} }

@article{Mozolin2000,
   Author = {Mozolin, Mikhail and Thill, J-C and Usery, E Lynn},
   Title = {Trip distribution forecasting with multilayer perceptron neural networks: A critical evaluation},
   Journal = {Transportation Research Part B: Methodological},
   Volume = {34},
   Number = {1},
   Pages = {53-73},
      Year = {2000}}

@article{Dong2018,
   Author = {Dong, Chunjiao and Shao, Chunfu and Clarke, David B and Nambisan, Shashi S},
   Title = {An innovative approach for traffic crash estimation and prediction on accommodating unobserved heterogeneities},
   Journal = {Transportation research part B: methodological},
   Volume = {118},
   Pages = {407-428},
      Year = {2018}}

@article{Cybenko1989,
   Author = {Cybenko, George},
   Title = {Approximation by superpositions of a sigmoidal function},
   Journal = {Mathematics of control, signals and systems},
   Volume = {2},
   Number = {4},
   Pages = {303-314},
      Year = {1989} }

@article{Hornik1991,
   Author = {Hornik, Kurt},
   Title = {Approximation capabilities of multilayer feedforward networks},
   Journal = {Neural networks},
   Volume = {4},
   Number = {2},
   Pages = {251-257},
      Year = {1991}}

@article{Poggio2017, 
   Author = {Poggio, Tomaso and Mhaskar, Hrushikesh and Rosasco, Lorenzo and Miranda, Brando and Liao, Qianli},
   Title = {Why and when can deep-but not shallow-networks avoid the curse of dimensionality: a review},
   Journal = {International Journal of Automation and Computing},
   Volume = {14},
   Number = {5},
   Pages = {503-519},
      Year = {2017} }

@article{Rolnick2017, 
   Author = {Rolnick, David and Tegmark, Max},
   Title = {The power of deeper networks for expressing natural functions},
   Journal = {arXiv preprint arXiv:1705.05502},
      Year = {2017} }

@incollection{Bousquet2004,
   Author = {Bousquet, Olivier and Boucheron, Stéphane and Lugosi, Gábor},
   Title = {Introduction to statistical learning theory},
   BookTitle = {Advanced lectures on machine learning},
   Publisher = {Springer},
   Pages = {169-207},
      Year = {2004} }

@book{Ledoux2013,
   Author = {Ledoux, Michel and Talagrand, Michel},
   Title = {Probability in Banach Spaces: isoperimetry and processes},
   Publisher = {Springer Science & Business Media},
      Year = {2013} }

@article{Baehrens2010, 
   Author = {Baehrens, David and Schroeter, Timon and Harmeling, Stefan and Kawanabe, Motoaki and Hansen, Katja and MÃžller, Klaus-Robert},
   Title = {How to explain individual classification decisions},
   Journal = {Journal of Machine Learning Research},
   Volume = {11},
   Number = {Jun},
   Pages = {1803-1831},
      Year = {2010} }

@inproceedings{Ross2018,
   Author = {Ross, Andrew Slavin and Doshi-Velez, Finale},
   Title = {Improving the adversarial robustness and interpretability of deep neural networks by regularizing their input gradients},
   BookTitle = {Thirty-second AAAI conference on artificial intelligence},
      Year = {2018}}

@article{Szegedy2014,
   Author = {Szegedy, Christian and Zaremba, Wojciech and Sutskever, Ilya and Bruna, Joan and Erhan, Dumitru and Goodfellow, Ian and Fergus, Rob},
   Title = {Intriguing properties of neural networks},
   Journal = {arXiv preprint arXiv:1312.6199},
      Year = {2014}}

@article{Rao1998,
   Author = {Rao, PV Subba and Sikdar, PK and Rao, KV Krishna and Dhingra, SL},
   Title = {Another insight into artificial neural networks through behavioural analysis of access mode choice},
   Journal = {Computers, environment and urban systems},
   Volume = {22},
   Number = {5},
   Pages = {485-496},
      Year = {1998} }

@article{Hensher2000,
   Author = {Hensher, David A and Ton, Tu T},
   Title = {A comparison of the predictive potential of artificial neural networks and nested logit models for commuter mode choice},
   Journal = {Transportation Research Part E: Logistics and Transportation Review},
   Volume = {36},
   Number = {3},
   Pages = {155-172},
      Year = {2000} }

@article{Allahviranloo2013,
   Author = {Allahviranloo, Mahdieh and Recker, Will},
   Title = {Daily activity pattern recognition by using support vector machines with multiple classes},
   Journal = {Transportation Research Part B: Methodological},
   Volume = {58},
   Pages = {16-43},
      Year = {2013} }

@article{Tang2015,
   Author = {Tang, Liang and Xiong, Chenfeng and Zhang, Lei},
   Title = {Decision tree method for modeling travel mode switching in a dynamic behavioral process},
   Journal = {Transportation Planning and Technology},
   Volume = {38},
   Number = {8},
   Pages = {833-850},
      Year = {2015}}

@article{ChengLong2019,
   Author = {Cheng, Long and Chen, Xuewu and De Vos, Jonas and Lai, Xinjun and Witlox, Frank},
   Title = {Applying a random forest method approach to model travel mode choice behavior},
   Journal = {Travel behaviour and society},
   Volume = {14},
   Pages = {1-10},
      Year = {2019}}

@article{Breiman2001,
   Author = {Breiman, Leo},
   Title = {Statistical modeling: The two cultures (with comments and a rejoinder by the author)},
   Journal = {Statistical science},
   Volume = {16},
   Number = {3},
   Pages = {199-231},
      Year = {2001} }

@article{Bertsimas2019,
   Author = {Bertsimas, Bimitris and Delarue, Arthur and Jaillet, Patrick and Martin, Sebastien},
   Title = {The Price of Interpretability},
   Journal = {Arxiv preprint},
      Year = {2019} }
\end{document}